\documentstyle{article}

\def\NCB{Il Nuovo Cimento {\bf B}}

\def\JMP{J.Math.Phys.}
\def\IM{Indag. Math. {\bf A}}

\def\PRD{ Phys. Rev. {\bf D}}

\def\CMP{ Commun. Math. Phys.}

\newtheorem{Th}{Theorem}
\newtheorem{Def}{Definition}
\newtheorem{emrem}{Remark}

{
\newtheorem{emexa}{Example}

\newtheorem{Prop}{Proposition}
\newtheorem{Cor}{Corollary}
\newtheorem{Lem}{Lemma}
{

 \def\pd#1,#2{\frac{\partial#1}{\partial#2}}
 \def\lpd#1,#2{\frac{\stackrel{\rightarrow}{\partial#1}}{\partial#2}}
\def\rpd#1,#2{\frac{\stackrel{\leftarrow}{\partial#1}}{\partial#2}}

\def\ra{\rightarrow}

\def\be{\begin{equation}}
\def\ee{\end{equation}}
\def\bea{\begin{eqnarray}}
\def\eea{\end{eqnarray}}
\def\nn{\nonumber}

\begin{document}
\thispagestyle{empty}
\vspace{.5cm}
\begin{center}
\vspace{1cm}

{\LARGE\bf The local structure\\[2mm] of $n$-Poisson and $n$-Jacobi manifolds} 
\footnote[1]
{Supported in part by the italian Ministero
dell' Universit\`a e della Ricerca Scientifica e Tecnologica.}
\end{center}
\medskip
\centerline{by}
\medskip

\begin{center}
{\bf G. Marmo} $^{1}$, {\bf G. Vilasi} $^{2}$, {\bf A.M.Vinogradov} $^{3}$
\end{center}
\bigskip
\begin{center}

$^{1}${\it Dipartimento di Scienze Fisiche , Universit\`a di Napoli,\\
Istituto Nazionale di Fisica Nucleare, Sezione di Napoli, Italy.}
\smallskip

$^{2}${\it Dipartimento di Scienze Fisiche} {\sl E.R.Caianiello}, 
{\it Universit\`a di Salerno,\\
Istituto Nazionale di Fisica Nucleare, Gruppo Collegato di Salerno, Italy.}

$^{3}${\it Dipartimento di Ing. informatica e Matematica Appl., 
Universit\`a di Salerno,\\
Istituto Nazionale di Fisica Nucleare, Gruppo Collegato di Salerno, Italy.}
\end{center}

\begin{abstract}
$n$-Lie algebra structures on smooth function algebras  
given by means of multi-differential operators, are studied.
\par
Necessary and sufficient conditions for the sum and the
wedge product of two $n$-Poisson sructures to be again a multi-Poisson are found. 
It is proven that the canonical $n$-vector on the dual of an $n$-Lie algebra 
$g$ is $n$-Poisson iff $dim~g\le n+1$. 
\par
The problem of compatibility of two $n$-Lie algebra
structures is analyzed and the compatibility relations connecting  hereditary
structures of a given $n$-Lie algebra are obtained. 
($n+1$)-dimensional $n$-Lie algebras are classified and their 
"elementary particle-like" structure 
is discovered.
\par
Some simple applications to dynamics are discussed.
\end{abstract}
\vskip 2truecm
\footnotesize
{\it Keywords: $n$-Lie algebra, $n$-Poisson (Nambu) bracket, $n$-Poisson 
(Nambu) manifold, $n$-Jacobi manifold.
}
\normalsize
\newpage
\tableofcontents
\vfill
\newpage
\section{Introduction}

 The concept of $n$-Poisson structure (Nambu-Poisson manifold in 
terminology by Takhtajan) is a particular case of that of $n$-Lie algebra. 
To our knowledge the latter was introduced for the fist time by V.T.Filippov 
\cite{Fi85} in 1985 who gave first examples, developed first 
structural concepts, like simplicity, in this context and  classified 
$n$-Lie algebras of dimensions $2n+1$ which is parallel to the  Bianchi 
classification of 3-dimensional Lie algebras. Filippov defines an $n$-Lie 
algebra structure to be an $n$-ary multi-linear  and anti-symmetric 
operation which satisfies the {\it $n$-ary Jacobi identity} : 

\bea\label{JI}
[[u_1,...,u_n],v_1,...,v_{n-1}]]&=&[[u_1,v_1,...,v_{n-1}],u_2,...,u_n]\cr
&+&[u_1,[u_2,v_1,...,v_{n-1}],u_3,...,u_n]+\cr 
&~&\cdots\cr
&+&[u_1,...,u_{n-1},[u_n,v_1,...,v_{n-1}]]
\eea

Such an operation, realized on the smooth function algebra of a manifold and 
additionally assumed to be an $n$-derivation, is an {\it $n$-Poisson structure}. 
This general concept, however, was not introduced neither by Filippov, nor, 
to our knowledge, by other mathematicians that time. It was done much later 
in 1994 by L.Takhtajan \cite{Ta94} in order to formalize mathematically 
the $n$-ary generalization of Hamiltonian mechanics proposed by Y.Nambu 
\cite{Na73} in 1973. Apparently Nambu was motivated by some problems of quark 
dynamics and the $n$-bracket operation he considered was :
\be\label{DET}          
\{f_1,...,f_n\} = det \|{\partial f_i\over \partial x_j}\|    
\ee
But Nambu himself as well as his followers do not mention that $n$-bracket 
($\ref{DET}$) satisfies the $n$-Jacobi identity ($\ref{JI}$). 
On the other hand, Filippov  reports ($\ref{DET}$) in his paper among 
other  examples of $n$-Lie algebras. It seems that Filippov's work remained 
unnoticed by physicists. For instance, Takhtajan refers in \cite{Ta94}  
to a private communication by Flato and Fronsdal of 1992 who observed that 
the Nambu {\it canonical} bracket ($\ref{DET}$) satisfies the 
{\it fundamental identity} ($\ref{JI}$). 
\par
  In this paper we study local $n$-Lie algebras, i.e. $n$-Lie algebra 
structures on smooth function algebras of smooth manifolds which are 
given by means of multi-differential operators. It follows from a theorem 
by Kirillov that these structure multi-differential operators are 
of first order. We call  {\it $n$-Jacobi} a local $n$-Lie algebra structure on 
a manifold. In the case when the structure multi-differential operator 
is a multi-derivation one gets an $n$-Poisson structure. So, $n$-Poisson 
manifolds form a subclass of $n$-Jacobian ones. The main mathematical result 
of the paper is a full local description of $n$-Jacobi and, in particular, 
of $n$-Poisson manifolds. This is an $n$-ary analogue of the {\it Darboux lemma}. 
In what concerns $n$-Poisson manifolds the same result was also recently  
obtained by Alexeevsky and Guha \cite{AG96}. Our approach is, however, quite 
different and, maybe, better reveals  why $n$-Poisson and $n$-Jacobi structures 
reduce essentially to the functional determinants ($\ref{DET}$) 
(theorems $\ref{Th1}$ and $\ref{Th2}$). 
\par   
An important consequence of the $n$-Darboux lemma is that the cartesian product 
of two $n$-Jacobi, or two $n$-Poisson manifolds does not produce manifold of 
the same type if $n>2$. 
Possibly this fact may explain the remarkable inseparability of quarks. 
This possibility suggests to investigate better the relevance of local $n$-Lie 
algebra structures for quark dynamics. The structure of 
($n+1$)-dimensional $n$-Lie algebras which is described in sect. 6 seems to be in 
favor of such idea.
\par   
It was not our unique goal in this paper to describe local structure of local $n$-Lie 
algebras. First, we tried to be systematic in what concerns the relevant basic 
formulae and 
constructions. Second, possible applications of the developed theory to integrable 
systems 
and related problems of dynamics are illustrated on some examples of current 
interest. 
\par			
More precisely, the content of the paper is as follows.
\par   
In sect. 2 the necessary generalities concerning $n$-Lie algebras and their 
derivations are 
reported. A new point discussed there is the concept of {\it compatibility} 
of two $n$-Lie structures 
defined on the same vector space. Two compatible structures can be combined 
to get a third one. This is why this concept seems to be of a crucial importance 
even for the theory of usual, i.e. 2-Lie, algebras. Fixing a number of arguments in an 
$n$-Lie bracket one gets new multi-linear Lie algebras of lower multiplicities, called 
{\it hereditary}. We deduce the compatibility relations tacking together 
hereditary structures of 
a given $n$-Lie algebra.
\par   
Generalities on $n$-Poisson manifolds are collected in sect. 3. 
There we introduce and discuss such basic notions related to an $n$-Poisson 
manifold as the {\it Casimir algebra},  {\it Casimir map} and 
{\it Hamiltonian foliation}. It is shown that 
an $n$-Poisson structures allow for multiplication by smooth functions 
if $n\geq 3$.
\par   
The main structure result regarding $n$-Poisson structures (theorem $\ref{Th1}$) 
is proved in sect. 4. 
It tells that the structure $n$-vector of an $n$-Poisson structure is of rank $n$ 
({\it decomposable}) if $n>2$. This leads directly to the {\it $n$-Darboux lemma}: 
{\it Given an $n$-Poisson structure, $n>2$, on a manifold $M$ there exists a local chart 
$x_1,...,x_m,~~ m=dim M \geq n$, on $M$ such that the corresponding $n$-Poisson bracket 
is given by ($\ref{DET}$)}. Two consequences of this result 
are worth  mentioning. First,the $n$-bracket defined naturally on 
the dual of an $n$-Lie algebra ${\cal V}$ is not generally an $n$-Poisson structure if $n>2$. 
This is in sharp contrast with usual, i.e. $n=2$, Lie algebras. However, we show that it is 
still so for $n$-dimensional and ($n+1$)-dimensional $n$-Lie algebras. 
By this and some other reasons it is naturally to conjecture that $n$-Lie algebras 
with $n>2$ are essentially $n$-dimensional and ($n+1$)-dimensional ones. 
Finally, in this section we deduce necessary and sufficient conditions in order the 
wedge product of two multi-Poisson structures be again a Poisson one.
\par   
The {\it $n$-Darboux} lemma for general $n$-Jacobi manifolds with $n>2$ is proved 
in sect. 5, theorem $\ref{Th2}$  and corollary $\ref{C13}$. 
The key idea in doing that is to split a first order multi-differential operator 
into two parts similarly to the canonical representation of a scalar first order 
differential operator as the sum of a derivation and a function. 
   An $n$-ary analogue of the well-known Bianchi classification of 3-dimensional Lie is 
given in sect. 6. An exhaustive description of ($n+1$)-dimensional $n$-Lie 
algebras was already done by Filippov \cite{Fi85} by a direct algebraic approach. 
Our approach is absolutely different and based on the use of the natural $n$-Poisson 
structure on the dual of an ($n+1$)-dimensional $n$-Lie algebra. 
It allows to get the classification in a very simple and transparent way and, what 
is more important, to discover what we would like to call a {\it elementary particle-like} 
structure of ($n+1$)-dimensional 
$n$-Lie algebras. More exactly, we shows that any such algebra is a specific linear 
combination of two {\it simplest} $n$-Lie algebra types realized in a mutually compatible 
(in the sense of sect. 2) way. A number similar to the coupling constant appears 
in this context. In this section we describe also derivations of ($n+1$)-dimensional 
$n$-Lie algebras and realize the Witt (or $sl(2,R)$-Kac-Moody) algebra as a 
2-Lie subalgebra of the canonical 3-algebra structure on ${\bf R}^3$.
   In the concluding sect. 7 we exhibit on concrete examples some simple applications of $n$-ary 
structures to dynamics. First, we use the Kepler dynamics to show how the constants of
motion can be put in relation with multi-Poisson structures. Second, alternative
Poisson realizations of a spinning particle dynamics $\Gamma$ are given by using 
ternary structures preserved by $\Gamma$. In a separate paper applications to dynamics
of the developed formalism will be discussed more systematically.
\par
   The multi-generalization of the concept of (local) Lie algebra studied in this paper is 
not, in fact, unique and there are other natural alternatives (see \cite{Mich Vin,Loday,Gned,VV}). 
All these generalizations are mutually interrelated and 
open very promising perspectives for particle and field dynamics.
\par
In this article we follow Filippov in what concerns the terminology and use 
{\it $n$-Lie algebra} instead of Takhtajian's {\it Nambu-Lie gebras}. 
The reason is  that arabic {\it al-gebre} became ethymologically indivisible in 
the current mathematical language , like {\it ring}, {\it group}, etc. 
So, it would be hardly convenient to use {\it $n$-gebra} together with 
indisputable {\it $n$-ring}.

\section{$n$-Lie algebras}
We start with some basic definitions.
\begin{Def}\label{D1} 
An $n$-Lie algebra structure on a vector space ${\cal V}$ (over a field $\bf K$) 
is a multi-linear mapping of 
$\underbrace{{\cal V}\times\cdots\times {\cal V}}_{n~~times}$ 
to ${\cal V}$ 
such that for any $u_i,v_j\in {\cal V}$, the $n$-Jacobi identity ($\ref{JI}$) holds.
\end{Def}
\begin{emrem}\label{R1} It is convenient to treat the ground field $\bf K$
 as the unique $0$-Lie algebra and a linear space supplied with a 
linear operator as an $1$-Lie algebra.
\end{emrem}
If an $n$-Lie algebra is fixed in the current context we refer to 
the underlying vector space ${\cal V}$ as the $n$-Lie algebra in question 
(as it is common for the {\it usual} Lie algebras). However, sometimes 
we need consider two or more $n$-Lie algebras structures on the same 
vector space. 
In such a situation we use $P(u_1,...,u_n)$ instead  of $[u_1,...,u_n]$. 
This notation appeals directly to the $n$-Lie algebra in question and 
is more flexible than the use
of alternative bracket graphics.

\begin{emexa}\label{E1}{\rm \cite{Fi85}} 
Let ${\cal V}$ be an $(n+1)$-dimensional vector space over $\bf R$  supplied
 with an orientation and a scalar product $(\cdot~,~\cdot)$. 

The {\it $n$-vector product} $[v_1,\dots,v_n]$ of 
$\;v_1,\dots,v_n\in {\cal V}\;$ is defined uniquely by requirements:
\begin{enumerate}
\item $[v_1,\dots,v_n]$ is ortogonal to all $v_i$'s;
\item $|[v_1,\dots,v_n]|=det\Vert (v_i,v_j)\Vert ^{\frac{1}{2}};$
\item the ordered system $v_1,\dots,v_n,[v_1,\dots,v_n]$ conforms the 
orientation of ${\cal V}$.
\end{enumerate}
\end{emexa}
Let $P$ and $Q$ be $n$-Lie algebra structures on ${\cal V}$ and ${\cal W}$, 
respectively. 
Then their direct product $R=P\oplus Q$ defined as
$$
R((v_1,w_1),\dots,(v_n,w_n))=(P(v_1,\dots,v_n),Q(w_1,\dots,w_n))
$$
 with $\;v_i\in {\cal V}, \;w_i\in {\cal W}\;$ is an $n$-Lie algebra structure
 on ${\cal V}\oplus {\cal W}$.

A central notion in the theory of $n$-Lie algebras is that of derivation \cite{Fi85}.
   
\begin{Def}\label{D2} 
A linear map ${\bf\cal D}:{\cal V}\rightarrow {\cal V}$ is said to be a 
derivation of the $n$-Lie 
algebra ${\cal V}$ if for any $u_1,...,u_n\in {\cal V}$ 
\be\label{d2}       
 {\bf\cal D}[u_1,...,u_n]= \sum_{i=1}[u_1,...,{\bf\cal D}u_i,...,u_n]     
\ee
\end{Def} 
 Fixing arbitrary elements $u_1,...,u_{n-1}\in {\cal V}$ one gets a map
$v\ra [u_1,...u_{n-1},v]$
which is a derivation of ${\cal V}$ as it follows from the Jacobi 
identity ($\ref{JI}$). 
Such a derivation is called {\it pure inner associated with} 
$u_1,...,u_{n-1}$. It will be denoted by $ad_{u_1,...,u_{n-1}}$ or 
$P_{u_1,...,u_{n-1}}$ for the $n$-Lie algebra structure $P$ in question. 
Linear combinations of pure inner derivations will be called 
{\it inner derivations (of $P$)}. 
Note that the concepts of inner and pure inner coincide for $n=2$ and that 
Hamiltonian vector fields are inner derivations of the background 
Poisson structure.
Following the standard terminology we, sometimes, shall call {\it outer}, 
derivations of ${\cal V}$ which are not inner  just to stress the instance of it.

\begin{Prop}\label{P1} 
Derivations of an $n$-Lie algebra form a Lie algebra with respect 
to the standard commutation operation and inner derivations constitute 
an ideal of it.
\end{Prop}
{\it Proof.} Let ${\bf\cal D}_1,{\bf\cal D}_2$ be derivations of the bracket 
$[\cdot ,\dots ,\cdot]$. Then, obviously,
\bea
{\bf\cal D}_1({\bf\cal D}_2([u_1,...,u_n])) = 
\sum_{i<j}([...,{\bf\cal D}_1u_i,...,{\bf\cal D}_2u_j,...]\cr+
[...,{\bf\cal D}_2u_i,...,{\bf\cal D}_1u_j,...]) +
\sum_i [u_1,...,{\bf\cal D}_1{\bf\cal D}_2u_i,...,u_n]
\eea

Therefore,
\be\label{e1}
[{\bf\cal D}_1,{\bf\cal D}_2]([u_1,...,u_n]) =
 \sum_i [u_1,...,[{\bf\cal D}_1,{\bf\cal D}_2]u_i,...,u_n]      
\ee
First assertion in the proposition is so proven. The second assertion follows
by observing that for a derivation ${\bf\cal D}$:
\bea\nonumber
[{\bf\cal D},ad_{u_1,...u_{n-1}}]u & =& {\bf\cal D}([u_1,...,u_{n-1},u])-
[u_1,...,u_{n-1},{\bf\cal D}u]\cr&=&\sum_{i\leq n-1} [u_1,...,{\bf\cal D}u_i,...u_{n-1},u]
\eea
So, in virtue of  ($\ref{e1}$) one has
\bea
&&[{\bf\cal D},ad_{u_1,...u_{n-1}}]([v_1,...,v_n])= 
\sum_i [v_1,...,[{\bf\cal D},ad_{u_1,...,u_{n-1}}]v_i,...,v_n]\cr
&&=\sum_i (\sum_{s\leq n-1}[v_1,\dots ,[u_1,...,{\bf\cal D}u_s,...u_{n-1},v_i],\dots ,v_n])\cr 
&&=\sum_{s\leq n-1} (\sum_i [v_1,\dots ,[u_1,...,{\bf\cal D}u_s,...u_{n-1},v_i],\dots ,v_n])\cr
&&=\sum_{s\leq n-1} ad_{u_1,...,{\bf\cal D}u_s,..,u_{n-1}}([v_1,...,v_n])
\eea

In other words,
\be\label{e2}
[{\bf\cal D},ad_{u_1,...u_{n-1}}] = 
\sum_{s\leq n-1} ad_{u_1,...,{\bf\cal D}u_s,..,u_{n-1}}       
\ee
or, with the alternative notation
\be\label{e3}
[{\bf\cal D},P_{u_1,...u_{n-1}}] = 
\sum_{s\le n-1} P_{u_1,...,{\bf\cal D}u_s,..,u_{n-1}}         
\ee 
\quad $\triangleright$

By putting ${\bf\cal D} = P_{v_1,...v_{n-1}}$ in ($\ref{e3}$) one gets the 
commutation formula 
for pure inner derivations :
\be\label{e5}
[P_{v_1,...v_{n-1}},P_{u_1,...u_{n-1}}] = \sum_i P_{u_1,...,[v_1,...v_{n-1},u_i],...,u_{n-1}}      
\ee
Note also the following relation in the algebra of inner derivations of $P$
which is due to skew-commutativity of the left hand side commutator in 
($\ref{e5}$):

$\sum_i P_{u_1,...,[v_1,...v_{n-1},u_i],...,u_{n-1}} +
\sum_i P_{v_1,...,[u_1,...,u_{n-1},v_i],...,v_{n-1}} = 0$
\par
  A description of the derivation algebra of an $n+1$-dimensional
 $n$-Lie algebra is given in proposition $\ref{DER}$, see also \cite{Fi85}.
Various outer derivations of an "atomic" 4-dimensional 3-lie algebra
are presented in example $\ref{E9+1}$.
\par
While the above results are just straightforward generalizations of 
known el\-emen\-tary facts of the standard Lie algebra theory the 
following simple observation (due to Filippov) is a very important 
new peculiarity of $n$-ary Lie algebras with $n>2$.

\begin{Prop}\label{P2}
 Let $P$ be an $n$-Lie al\-ge\-bra struc\-ture on ${\cal V}$. 
Then for any
$u_1,...,u_k\in {\cal V},~~ k\leq n$, 
$P_{u_1,...,u_k}$ is an $(n-k)$-Lie algebra 
structure on ${\cal V}$.
\end{Prop}
 {\it Proof.} It is sufficient, obviously, to prove this result for $k=1$ only. 
But in this case one can see easily that the Jacobi identity for $P_u$
is obtained from that of $P$ just by putting in it $u_n=u_{n-1}=u$. $\triangleright$

\begin{emexa}\label{E2} 
If $P$ is the $n$-vector product structure of example $\ref{E1}$, then the 
$(n-k)$-Lie algebra structure $P_{u_1,\dots,u_k}$ on ${\cal V}$ is the direct product of
the trivial structure on $S= Span\{u_1,\dots,u_k\}$ and the $(n-k)$-vector product
structure on $S^\perp$ with respect to the scalar product
$$
(\cdot,\cdot)^\prime=\lambda(\cdot,\cdot)|_{S^\perp}\quad,
\quad \lambda=(vol_k(u_1,\dots,u_k))^{\frac{1}{n-k}},
$$
on $S^\perp$.
\end{emexa}
   Multi-Lie structures $P_{u_1,...,u_k}$ obtained in this way from $P$ 
will be called {\it hereditary} (with respect to $P$) of order $k$. 
The fact that these structures belong to the same {\it family} implies 
mutual {\it compatibility} of them, an important concept we are going 
to discuss.

   With this purpose we need first the following analogue of the 
Lie derivation operator. 
Let $Q :{\cal V} \times...\times {\cal V}\ra {\cal V}$ be a $k$-linear 
mapping and 
$\partial :{\cal V}\ra {\cal V}$ be a linear operator. 
The ${\partial}$-derivative ${\partial}(Q)$ of $Q$ is also a 
$k$-linear map defined as

$
[{\partial}(Q)](u_1,...,u_k) = {\partial}(Q(u_1,...,u_k)) - 
\sum_i Q(u_1,...,{\partial}u_i,...,u_k)
$

Note that the Jacobi identity ($\ref{JI}$) is equivalent to
$P_{u_1,...,u_k} (P) = 0$    for any $u_1,...,u_k\in {\cal V}$.
   
\begin{emexa}\label{E3} If $k = 1$, i.e. $Q$ is a linear operator on ${\cal V}$, then 
$\partial(Q) = [\partial,Q]$.
\end{emexa}   
Sometimes it is more convenient to use $L_{\partial}$ instead of $\partial$ 
for the $\partial$-derivative. An instance of it is the formula
\be\label{e6}
[L_{\partial},\imath_u] = \imath_{\partial(u)}                            
\ee
where $\imath_u$ for $u\in {\cal V}$ denotes the insertion operator, i.e.
\be
\imath_u(Q)(u_1,...,u_{k-1}) = Q(u,u_1,...,u_{k-1})            
\ee
The proof of ($\ref{e6}$) is trivial.
   
\begin{Def}\label{D3} Two $n$-Lie algebra structures on ${\cal V}$ are said compatible if
for any $u_1,...,u_{n-1}\in {\cal V}$.
\be\label{e7}
P_{u_1,...,u_{n-1}}(Q) +  Q_{u_1,...,u_{n-1}}(P) = 0     
\ee
\end{Def}
\begin{emrem}\label{R2} If ${\cal V} = C^\infty(M), n=2$ and $P$ and $Q$ are two 
Poisson structures on $M$, then they are compatible in the well-known 
sense of Magri \cite{Ma78}(see also \cite{DMSV82,DMSV84,LMV94}) iff they are compatible 
in the sense of definition $\ref{D3}$. It is not difficult to see that in such a situation 
condition ($\ref{e7}$) is identical to vanishing of the Schouten 
bracket of $P$ and $Q$.
\end{emrem}
\begin{emexa}\label{E4} For $n = 1$ the compatibility condition is empty. 
In fact, in this case $P$ and $Q$ are just linear operators and 
$$P(Q)+Q(P)=[P,Q] +[Q,P]= 0$$.
\end{emexa}
  The following proposition gives a possible interpretation of the 
notion of compatibility.
   
\begin{Prop}\label{P3}
 Let $P$ and $Q$ be $n$-Lie structures on ${\cal V}$. 
If $a,b\in {\bf K},~~ab\neq 0$, then  $aP+bQ$ is an $n$-Lie algebra structure 
iff $P$ and $Q$ are compatible.
\end{Prop}   
{\it Proof.}
The following identity is due to linearity of the {\it Lie derivative}
expression $I(R)$
 with respect to both $I$ and $R$:
\bea\nn
(aP+bQ)_{u_1,...,u_{n-1}}(aP+bQ) &= &a^2 P_{u_1,...,u_{n-1}}(P)+  
abP_{u_1,...,u_{n-1}}(Q)\cr 
&+&abQ_{u_1,...,u_{n-1}}(P) + b^2 Q_{u_1,...,u_{n-1}}(Q)
\eea
It remains now to apply interpretation ($\ref{e6}$) of the Jacobi identity.$\triangleright$ 

\begin{emexa}
Let {\cal A} be an associative algebra. For a given $M\in {\cal A}$ 
define a skew-symmetric bracket $[\cdot~,~\cdot ]_M$ on ${\cal A}$
by putting 
\begin{equation}
[A,B]_M=AMB-BMA,~~~~A,B\in {\cal A}.
\end{equation}\nonumber

It is easy to see that this, in fact, is a Lie algebra structure on
${\cal A}$. Moreover, for any $M,N\in {\cal A}$ structures
$[\cdot~,~\cdot ]_M$ and $[\cdot~,~\cdot ]_N$ are compatible. This
follows from the fact that 
$$
[\cdot~,~\cdot ]_M~ + ~[\cdot~,~\cdot ]_N~ = ~[\cdot~,~\cdot ]_{M+N} 
$$
\end{emexa}

\begin{Cor}\label{C1} Any two first order  hereditary structures $P_u$ and $P_v$
of an $n$-Lie algebra $P$ are compatible.
\end{Cor}  
{\it Proof.}~~In fact, according to proposition $\ref{P2}$,  $P_u+P_v = P_{u+v}$ is an 
$(n-1)$-algebra structure. $\triangleright$
     
On the contrary, hereditary structures of order greater than $1$ are not, 
in general, mutually compatible . It can be seen as follows.
   
Denote by $Comp(P,Q;u_1,...,u_{n-1})$ the left hand side of the compatibility 
condition ($\ref{e7}$). Then a direct computation shows that
\bea\nn
Comp(P_{u,v},P_{w,z};u_1,..,u_{n-3})&=& P_{P(u,v,u_1,..,u_{n-3},w),z} + 
P_{w,P(u,v,u_1,..,u_{n-3},z)}\cr 
&+& P_{P(w,z,u_1,..,u_{n-3},u),v}+P_{u,P(w,z,u_1,..,u_{n-3},v)}
\eea
In particular, for $u_1 = u$ we have
\bea\nn
Comp(P_{u,v},P_{w,z};u,u_2,...,u_{n-3})& =& P_{u,P(w,z,u,u_2...,u_{n-3},v)}\cr 
& =& Q_{Q(w,z,u_2...,u_{n-3},v)}
\eea
with $Q = P_u$. Now one can see from an example that 
$Q_{Q(w,z,u_2...,u_{n-3},v)}$ is generically different from zero. 
For instance, if $P$ is the $n$-vector product
algebra, then $Q = P_u$ is isomorphic to the direct sum of the 
$(n-1)$-vector product algebra and the trivial $1$-dimensional one. 
Then $Q_{Q(w,z,u_2...,u_{n-3},v)} = 0$ for linearly independent 
$w,z,u_2...,u_{n-3},v$ belonging to the first direct summand.
  However, second order hereditary structures are subjected to another 
kind of relations deriving from that of compatibility. 
To describe them it will be convenient to introduce a symmetric 
bilinear function $Comp(P,Q)$ defined by:
\be
Comp(P,Q)(u_1,...,u_{n-1}) = Comp(P,Q;u_1,...,u_{n-1})\nn
\ee
By definition $Comp(P,Q)$ is an $(n-1)$-linear skew-symmetric function 
on ${\cal V}$ with values in the space of $n$-linear skew-symmetric functions 
on ${\cal V}$. By this reason we have, in particular,
\bea\nn
 Comp(P_{u+w,v},P_{u+w,z}) &=& Comp(P_{u,v},P_{u,z}) + 
Comp(P_{u,v},P_{w,z})\cr
&+&Comp(P_{w,v},P_{u,z}) + Comp(P_{w,v},P_{w,z})  
\eea
Note now that two second order hereditary structures of the form 
$P_{x,y},~~P_{x,z}$ are compatible because they can be regarded as first 
order hereditary structures of the $(n-1)$-Lie algebra $P_x$. 
By this reason the above equality reduces to
\be\label{e8}
       Comp(P_{u,v},P_{w,z}) + Comp(P_{u,z},P_{w,v}) = 0            
\ee
Identity ($\ref{e8}$) binding second order secondary structures 
tells that the compatibility condition between $P_{u,v}$ and $P_{w,z}$ 
depends rather on bi-vectors $u\wedge v$ and $w\wedge z$ than 
on vectors $u,v$ and $w,z$ representing them, correspondingly.
\par   
Similar relations binding together $k$-th order hereditary structures 
can be found by generalizing properly the above reasoning. 
With this purpose we need to develop a suitable notation associated
with a fixed $n$-Lie algebra structure $P$ on ${\cal V}$. 
Let $v_1,...,v_k , w_1,...,w_k\in{\cal V},~~ i =1,...,k$. 
\par
Let us define the symbol $<v_1,...,v_k | w_1,...,w_k>$
by:

\bea\nonumber
&&<v_1,..,v_k | w_1,..,w_k>(u_1,..,u_{n-k-1})\cr 
&&=Comp(P_{v_1,..,v_k},P_{w_1,..,w_k};u_1,..,u_{n-k-1})
\eea

So, $<v_1,...,v_k | w_1,...,w_k>$ is a skew-symmetric 
$(n-k-1)$-linear function on ${\cal V}$
with values in the space of $(n-k)$-linear skew-symmetric functions on ${\cal V}$. 
Moreover, it is symmetric with respect $v$ and $w$, i.e.
\be\nn
<v_1,...,v_k | w_1,...,w_k> = <w_1,...,w_k | v_1,...,v_k>
\ee
and skew-symmetric with respect to variables $v_i$'s as well as $w_i$'s.
   If $I = (i_1,...,i_p)$ is a sequence of integers such that 
$i_1<...<i_p$, then $(v,w)_I$ stands for the sequence of $n$ elements of 
${\cal V}$ such that its $s$-th term
is $v_s$ if $s\in I$ and $w_s$ otherwise. 
A similar meaning has the symbol $(w,v)_I$. For example, if $k = 5$
 and $I = (1,3)$, then $(v,w)_I =(v_1,w_2,v_3,w_4,w_5)$,~~
$(w,v)_I= (w_1,v_2,w_3,v_4,v_5)$. Define now the following {\it quadratic} 
function :
\be\label{e9}
C(v_1,...,v_k | w_1,...,w_k)=\sum_{I,i_1=1}<(v,w)_I|(w,v)_I >        
\ee   

\begin{Prop}\label{P4} 
For any $v_1,...,v_k , w_1,...,w_k \in {\cal V}, n\geq k,$ it holds
\be\label{e10}
            C(v_1,...,v_k | w_1,...,w_k) = 0            
\ee    
\end{Prop}
Equality ($\ref{e10}$) is called {\it the $k$-th order compatibility condition}.

\begin{emrem}\label{R3} Corollary $\ref{C1}$ is identical to ($\ref{e10}$) for $k = 1$ 
while formula ($\ref{e8}$) to ($\ref{e10}$) for $k = 2$.
\end{emrem}   
{\it Proof.} It goes by induction. Corollary $\ref{C1}$ allows to start it. 
Supposing then the validity of ($\ref{e10}$) for $k$ for all 
multi-Lie algebras, we observe that
\be\nn 
C(x_1,...,x_k,u | y_1,...,y_k,u) = 0
\ee
(for any $x_1,...,x_k, y_1,...,y_k,u\in{\cal V}$. 
In fact,this condition coincides with the $k$-th order compatibility 
condition for $(n-1)$-Lie algebra $P_u$. In particular,
\be\nn
C(v_1,...,v_k, v_{k+1}+w_{k+1} | w_1,...,w_k, v_{k+1}+w_{k+1}) = 0
\ee
On the other hand, it is easily seen that
\bea\nn
&C(v_1,..,v_k, v_{k+1}+w_{k+1} | w_1,..,w_k, v_{k+1}+w_{k+1})=\cr
&\sum_{I,i_1=1} <(v,w)_I, v_{k+1}+w_{k+1}|(w,v)_I,v_{k+1}+w_{k+1}>
\eea
where $(v,w)_I$ has the same meaning as in ($\ref{e9}$) and $((v,w)_I,x)$ 
denotes the sequence that becomes $(v,w)_I$ once last term $x$ is deleted. 
Multi-linearity of the symbol $<...| ...>$ allows to develop 
last expression as the sum of terms of the form $<(v,w)_I,x|(w,v)_I,y>$
 with $x,y$ taking independently the values $v_{k+1},w_{k+1}$ . 
After that it remains to observe that the $k$-th order compatibility
condition for the algebra $P_x$ gives
\be\nn
\sum_{I,i_1=1} <(v,w)_I,x|(w,v)_I,x> = 0    
\ee\nn
and
\bea\nn
C(v_1,..., v_{k+1} | w_1,...,w_{k+1})& = &
\sum_{I,i_1=1} <(v,w)_I, v_{k+1}|(w,v)_I,w_{k+1}>\cr 
&+&\sum_{I,i_1=1} <(v,w)_I, w_{k+1}|(w,v)_I,v_{k+1}>
\eea $\triangleright$
  
\begin{emexa}\label{E5}
 The explicit form of the third compatibility condition is
\bea\nn
&Comp(P_{v_1,v_2,v_3},P_{w_1,w_2,w_3})+ Comp(P_{v_1,v_2,w_3},P_{w_1,w_2,v_3})\cr +
&Comp(P_{v_1,w_2,v_3},P_{w_1,v_2,w_3}) + Comp(P_{v_1,w_2,w_3},P_{w_1,v_2,v_3}) = 0.
\eea
\end{emexa}
The second order compatibility conditions provides some necessary conditions 
for the following natural question:
  
{\it Whether two given $n$-Lie algebra structures $Q$ and $R$ come from a 
common $(n+1)$-Lie algebra structure, i.e. whether $Q = P_u$, $R = P_v$ 
for an $(n+1)$-Lie algebra $P$ and some $u,v\in {\cal V}$ ?}
   
\begin{Cor}\label{C2} 
If $n$-Lie algebra structures $Q$ and $R$ are first order hereditary for 
an $(n+1)$-Lie algebra, then
\be
Comp(Q_w,R_z) + Comp(Q_z,R_w) = 0, ~~~~~\forall w,z\in {\cal V}.
\ee 
\end{Cor}
\section{$n$-Poisson manifolds}

   The concept of $n$-Poisson manifold generalize the one of Poisson one ($n=2$)
just in the same sense as $n$-Lie algebras do with respect to Lie algebras. 
It was introduced by Takhtajan in \cite{Ta94}. Filippov in his pioneering work \cite{Fi85} 
gives an example (see example $\ref{E7}$ below) which turned out to be locally 
equivalent to  the general concept in virtue of an analogue of the Darboux 
lemma for $n$-Poisson structures. This analogue was found recently by 
Alekseevsky and Guha \cite{AG96}. Below we present a simple purely algebraic 
proof of it which is valid in more general algebraic contexts, for 
instance, for smooth algebras. Since $n$-Poisson structures are special 
kind of $n$-Lie algebra ones we can use freely results of the preceding 
section in this context.
   
\begin{Def}\label{D4} Let $M$ be a smooth manifold. An $n$-Lie algebra 
structure on $C^\infty (M)$
\be                  
(f_1,...,f_n)\ra \{f_1,...,f_n\}\in C^\infty (M),~~~~f_i\in C^\infty (M)\nn 
\ee
is called an $n$-Poisson structure on $M$ if the map
\be
f\ra \{f,...,\}\nn
\ee
is a {\it derivation} of the algebra  $C^\infty (M)$.
\end{Def}  
Last condition means the Leibniz's rule with respect to the first argument :
\be
\{fg, h_1,...,h_{n-1}\} = f\{g, h_1,...,h_{n-1}\} + g\{f, h_1,...,h_{n-1}\}\nn 
\ee
Evidently, due to skew-symmetry, the Leibniz's rule is valid for all arguments.
    An equivalent way to express this property is to say that the operator
\be
X_{f_1,...,f_{n-1}} : C^\infty (M)\ra C^\infty (M)\nn 
\ee
defined as
\be
X_{f_1,...,f_{n-1}}(g) = \{f_1,...,f_{n-1},g\}\nn
\ee
is a vector field on $M$. Such a field is called {\it Hamiltonian} 
corresponding to the {\it Hamiltonian functions} $f_1,...,f_{n-1}$. 
   
A manifold supplied with an $n$-Poisson structure is called {\it $n$-Poisson 
or Nambu-Poisson manifold.}
   It is natural to interpret a vector field on $M$ as an $1$-Poisson 
structure on it.

Vector fields on $M$ that are derivations of the considered $n$-Poisson 
structure are called {\it canonical} (with respect to it). As in the 
classical case $n=2$ Hamiltonian fields of an $n$-Poisson structure
are, obviously,  canonical fields.
 
Let $M$ and $N$ be $n$-Poisson manifolds and $\{~ , ~\}_M$ and 
$\{~ , ~\}_N$ be 
the corresponding brackets. A map $F:M\ra N$ is said to be {\it Poisson} if
      
\be
\{F^*(f_1),...,F^*(f_n)\}_M = F^*(\{ f_1,...,f_n\}_N)\nn 
~~~\forall f_1,...,f_n\in C^\infty (M)
\ee   

\begin{emexa}\label{E6}{\rm\cite{Fi85}}
Let $X_1,...,X_n$ be commuting vector fields on $M$. 
Then
\be\label{CF}
\{f_1,...,f_n\} = det\|X_i(f_j)\|
\ee
is an $n$-Poisson structure on $M$. More generally, if ${\cal A}$ is a 
commutative algebra, any set of $n$ commuting derivations of it defines an 
$n$-Poisson structure on it. 
Note also that the so-defined $n$-Poisson structure is invariant with
respect to a unimodular transformation of fields $Y_i =\sum_js_{ij}X_j$,  
 $det\|s_{ij}\|=1, ~~~s_{ij}\in C^\infty(M)$.
\end{emexa}
More generally if $[~X_j,~X_k]=c^l_{jk}X_l,~~~c^l_{jk}\in C^\infty(M)$,
we have $\{f_1,...,f_n\} = det\|X_i(f_j)\|$ is an $n$-Poison structure on $M$.

   $n$-Poisson structures are multiderivations, i.e. multilinear 
operators on the algebra $C^\infty(M)$  which are derivations with respect to any of their 
arguments. This is a particular case of the general concept of 
multidifferential operator on $C^\infty(M)$ (more generally, on a commutative 
algebra ${\cal A}$ \cite{VinC}). It means that for any $i=1,2,...,k$ the 
correspondence
\be
f\ra \Delta(f_1,...f_{i-1},f,f_{i+1},...,f_k)\nn
\ee
is a differential operator for any fixed set of func\-tions 
$f_1,..,f_{i-1},f_{i+1},..,f_k$. 
When dealing with multidifferential operators and, 
in particular, with multiderivations we will adopt the notation of  
the previous section. For instance, we write $f\rfloor$  or $\imath_f$
 for the insertion operator. 
For instance, if $\Delta$ is a $k$-differential operator, 
then $f\rfloor \Delta = \imath_f(\Delta) = \Delta_f$ 
are three different notations for the $(k-1)$-differential operator
\be                      
(f\rfloor \Delta)(g_1,...g_{k-1}) = \Delta (f,g_1,...g_{k-1})\nn
\ee                                                                                                                                                                                                     

Note the one-to-one correspondence between $k$-contravariant tensors $T$ 
and $k$-derivations $\Delta$ 
given as
\be
df_k\rfloor ...\rfloor df_1\rfloor T = T(df_1,...,df_k) = \Delta(f_1,...,f_k)\nn
\ee
If, moreover, $T$ is skew symmetric, then it is a $k$-vector. 
In particular, an $n$-Poisson structure can be given either by a 
skew symmetric $n$-derivation, or by the $k$-vector corresponding to it . 
  
 The mentioned one-to-one correspondence between skew-symmetric 
multi-derivations and multi-vectors allows to carry well-known operations 
from the latters over the formers. For instance, the standard {\it wedge product} 
of two multi-vectors allows to define {\it the wedge product} of
the corresponding multiderivations $\Delta$ and $\nabla$ as 
\be\label{e11}
(\Delta\wedge \nabla)(f_1,...,f_{k+l}) = 
\sum_I (-1)^{(I,{\bar I})}\Delta (f_I)\nabla (f_{\bar I})     
\ee
where $I = (i_1,...,i_k),~~~ 1 \leq i_1\leq\dots\leq i_k\leq k+l$, is an increasing 
subsequence of integers, $\bar I$ is its complement in 
$\{1,2,...,k+l\}$,~~ $(I,\bar I)$ is the corresponding permutation of 
$1,2,...,k+l$,~~ $(-1)^{(I,{\bar I})}$ stands  for the sign of it and 
$f_I$ (respectively $f_{\bar I}$) is a shortnoting for $f_{i_1},...,f_{i_k}$ 
(respectively $f_{\bar{\imath}_1},...,f_{\bar{\imath}_l}$). 
Moreover, definition ($\ref{e11}$) makes sense, in fact, for arbitrary 
multi-differential operators, not necessarily derivation, and therefore, 
defines an associative and graded commutative multiplication over them.

The Schouten-Nijenhuis bracket carried over multi-derivations looks as
\bea\label{e12}
\lceil \Delta, \nabla \rfloor (f_1,...,f_{k+l-1})=\sum_{|I|=k-1} (-1)^{(I,{\bar I})}
\Delta(f_I, \nabla (f_{\bar I}))\cr -
\sum_{|J|=k} (-1)^{(J,{\bar J})}\nabla(\Delta (f_J),f_{\bar J})                     
\eea
where $I$ and $J$ stand, as before, for increasing subsequence of 
$\{1,2,...,k+l-1\}$ while $|I| $(respectively, $|J|$) denotes the length of $I$ 
(respectively, $J$).
Similarly to ($\ref{e11}$), formula ($\ref{e12}$) remains meaningful for arbitrary 
skew-symmetric
 multi-differential operators and this way the Schouten-Nijenhuis bracket 
is extended on them. More exactly, defining the {\it Schouten grading} of
 $k$-differential operators to be equal $k-1$, we have:

\begin{Prop}\label{P5} 
The {\it Schouten graded} skew-symmetric 
multi-differential operators supplied with the bracket operation 
($\ref{e12}$) form a graded Lie algebra, i.e.
\be           
\lceil \Delta,\nabla \rfloor = -(-1)^{(k-1)(l-1)}\lceil \nabla ,\Delta \rfloor\nn    
\ee 
(graded skew-symmetry)
and
\bea\nn     
&(-1)^{(k-1)(m-1)}\lceil \Delta ,\lceil \nabla,\Box \rfloor \rfloor +\cr
&(-1)^{(m-1)(l-1)}\lceil \Box, \lceil \Delta,\nabla \rfloor \rfloor +\cr
&(-1)^{(l-1)(k-1)}\lceil\nabla, \lceil \Box,\Delta \rfloor \rfloor =0
\eea                                                   
(graded Jacobi identity)
\end{Prop}

{\it Proof.} Graded skew-commutativity is obvious while the 
graded Jacobi identity is checked by a direct but tedious computation.$\quad\triangleright$ 
   
\begin{Cor}\label{C3} The well-known compatibility condition 
$ \lceil \Delta,\nabla \rfloor = O$ of two Poisson structures 
~~$\Delta(f,g) = \{f,g\}_I$ and $\nabla(f,g) = \{f,g\}_{II}$ is 
in the considered context identical to the one given in the preceding section.
\end{Cor}   
{\it Proof.} Just to compare ($\ref{e7}$) for $n = 2$ and ($\ref{e12})$ 
for $k=l=2$. 
   
\begin{emrem}\label{R4} It is worth to emphasize that the Lie derivative of a 
multi-vector  $V$ corresponds in the aforementioned sense to 
the {\it Lie derivative} in the sense of the previous section of 
multi-derivation $\Delta$  corresponding to $V$. 
In particular, the fact that $V$ is an $n$-Poisson multi-vector 
can be seen as
\be 
X_{f_1,...,f_{n-1}}(V) = 0\nn
\ee
where $X(V)$ is a short notation for the Lie derivative 
$L_X(V)$ of~~ $V$ we shall use to simplify some formulae. 
Similarly, the compatibility condition
of two $n$-vectors $V$ and $W$ can be written in the form
\be
Y_{f_1,...,f_{n-1}}(V) + X_{f_1,...,f_{n-1}}(W) = 0\nn
\ee
where  $X_{f_1,...,f_{n-1}}$ and $Y_{f_1,...,f_{n-1}}$ are 
Hamiltonian vector fields with 
the same {\it Hamilton functions} $f_1,...,f_{n-1}$ in the sense of Poisson 
structures given by  $V$ and $W$, respectively.
\end{emrem}
   A function $g\in C^\infty (M)$ is said to be a {\it Casimir function} 
if
\be
X_{f_1,...,f_{n-1}}(g) = \{f_1,...,f_{n-1},g\}=0, 
~~~\forall f_1,...,f_{n-1}\in C^\infty (M).\nn
\ee
 
All Casimir functions form, evidently, 
a subalgebra ${\cal K}$ of $C^\infty (M)$. 
We denote it also $Cas(P)$ when it becomes 
necessary to refer to the $n$-Poisson structure $P$ in question and call 
it the Casimir algebra .
   An ideal ${\cal I}$ of the Casimir algebra allows to restrict the original
 $n$-Poisson structure to the submanifold (possibly with singularities) 
\be                 
 N = \{ x\in M~~ |~~  f(x) = 0,~~~ f\in {\cal I}\}\subseteq M. 
\ee
To see this note that 
\be\label{e13}
C^\infty (N)=C^\infty (M)/{\cal I}C^\infty (M)              
\ee
if $N$ is a submanifold without singularities.
Otherwise, define the smooth function algebra on $N$ by means of 
($\ref{e13}$). Further note that the ideal 
${\cal I^{\ast}}={\cal I}C^\infty(M)\subseteq  C^\infty (M)$ 
is {\it stable} (with respect to the $n$-Poisson structure in question) 
in the sense that $\{f_1,...,f_{n-1},g\}\in {\cal I^{\ast}}$ if 
$g\in {\cal I^{\ast}}$. This allows one to 
define the restricted $n$-Poisson structure on $N$ just by passing 
to quotients
\be 
\{\widetilde{f}_1,...,\widetilde{f}_n\}_N = \widetilde{\{f_1,...,f_n\}}\nn
\ee 
where  $\widetilde{f}_i = f_i~~~(\bmod~{\cal I^{\ast}})$.
 From a geometrical point of view the stability of ${\cal I^{\ast}}$
implies that Hamiltonian vector fields are tangent to $N$.
 The smallest of such  submanifolds  $N$ correspond to the largest, i.e. 
maximal, ideals of ${\cal K}$. Since any {\it non wild} maximal ideal of 
${\cal K}$ is of 
the form ${\cal I} = ker~G$ where $G: {\cal K}\ra \bf R$  is a 
$\bf R$-homomorphism of  unitary 
$\bf R$-algebras it is reasonable to limit our considerations to these ones. 
Denote by $N_G$ the submanifold of $M$ corresponding to the ideal 
${\cal I} = kerG$ 
and recall that all $\bf R$-homomorphisms of ${\cal K}$ constitute a manifold 
(with singularities) $Spec_{\bf R}{\cal K}$, the real spectrum of ${\cal K}$, 
in such a way that 
${\cal K} = C^\infty(Spec_{\bf R}{\cal K})$. 
We shall call it the {\it Casimir manifold} of the considered 
$n$-Poisson structure and denote it by $Cas(M)$ or $Cas(P)$ depending of the 
context. Then the canonical embedding ${\cal K}\subseteq C^\infty(M)$ 
induces by duality the Casimir map
\be
Cas : M\ra Cas(M)\nn
\ee
By construction $N_G = Cas^{-1}(G)$. 
This way one gets the {\it Casimir fibration} of $M$ whose 
fibres are $n$-Poisson manifolds.
   In the Casimir fibration it is canonically inscribed the {\it Hamiltonian 
foliation} which is defined as follows. First, note that the commutator 
of two Hamiltonian fields is a sum of Hamiltonian fields. In fact, 
formula $\ref{e5}$ in the considered context looks as
\be
[X_{f_1,...,f_{n-1}},X_{g_1,...,g_{n-1}}] = 
\sum_iX_{g_1,...,\{f_1,...,f_{n-1},g_i\},...,g_{n-1}}
\ee
This implies that the $C^\infty(M)$-module $H(P)$ of vector fields 
generated by all Hamiltonian ones is closed with respect to the 
Lie commutator operation. It defines, there\-fore,  a (singular) foliation 
on $M$ called Hamiltonian.  It was already mentioned that Hamiltonian fields
 are tangent to submanifolds $N_G$ . Hence, any {\it Hamiltonian leaf}, i.e. 
that of the Hamiltonian foliation, belongs to a suitable Casimir 
submanifolds $N_G$. So, Casimir submanifolds are foliated by 
{\it Hamiltonian leaves}.
  
\begin{emexa}\label{E7} Let $T^{n+1}$ be the standard $(n+1)$-dimensional torus with 
standard angular coordinates 
$\theta_1, \theta_2,...,\theta_{n+1}$. 
Consider the $n$-Poisson 
structure on it defined by vector fields 
\be
X_1 = {\partial\over \partial\theta_1} + 
\lambda {\partial\over \partial \theta_2},~~  
X_2 = {\partial\over \partial \theta_3},..., 
X_n = {\partial\over \partial \theta_{n+1}}\nn 
\ee
as in example $\ref{E6}$. Then for a rational $\lambda$  $Cas(T^{n+1}) = S^1$ and the 
Casimir map  $Cas : T^{n+1}\ra  S^1$  is a trivial fibre  bundle with 
$T^n$ as 
fibre. In this case fibres of the Casimir map are identical to leaves 
of the Hamiltonian foliation. If  $\lambda$ is irrational, then $Cas(T^{n+1})$  
is just a point what is equivalent to ${\cal K} = {\bf R}$. In other words,  
$T^{n+1}$ is the unique submanifold of the form $N_G$. On the other hand, the 
Hamiltonian foliation in this case is $n$-dimensional and its leaves
are copies of ${\bf R}^n$  immersed everywhere densely in $T^{n+1}$.
\end{emexa}   
   Since Hamiltonian vector fields are, by construction, tangent to 
the leaves of the Hamiltonian foliation, the Poisson multi-vector of 
the considered Poisson structure is also tangent to them. For this reason 
on any such a leaf there exists an unique $n$-Poisson structure such that 
the canonical immersion $L\hookrightarrow M$ becomes an $n$-Poisson map. 
In the next 
section it will be shown that Poisson leaves are either $n$-dimensional 
(regular), or $0$-dimensional (singular) if  $n>2$ what is in a strong contrast 
with the classical case $n=2$. By this reason $n$-Poisson structures on 
$n$-dimensional manifolds are to be described. We will get it as a particular
case of the following general assertion.

\begin{Prop}\label{P6} 
 Let $P$ be an $n$-Poisson structure of rank $n$ on 
a manifold $M$.
Then for any $f\in {\cal C}^\infty(M)$,~$fP$ is an $n$-Poisson structure and 
any two structures of this form are compatible.
\end{Prop}
{\it Proof.} It is based on the general formula
\be
 L_{fX}(Q)=fL_X(Q)-X\wedge(f\rfloor Q)\nn
\ee
for any $f\in {\cal C}^\infty(M)~~,X\in{\cal D}(M)$
and a multi-vector $Q$ on $M$ (see, for instance,\cite{CV92}).
By applying it to $X=P_{h_1,\dots,h_{n-1}}$ 
 and $Q=gP,~~ g\in{\cal C}^\infty(M)$ and taking into account that 
$P_{h_1,\dots,h_{n-1}}(P)=0$ one finds
\be
(fP)_{h_1,\dots,h_{n-1}}(gP)=fP_{h_1,\dots,h_{n-1}}(g)P-P_{h_1,\dots,h_{n-1}}
\wedge(gP_f)\nn
\ee
This formula allows to rewrite the compatibility condition ($\ref{e7}$)
 for $fP$ and $gP$ as
\bea
&&(fP)_{h_1,\dots,h_{n-1}}(gP)+(gP)_{h_1,\dots,h_{n-1}}(fP)=\nn\cr
&&fP_{h_1,\dots,h_{n-1}}(g)P+gP_{h_1,\dots,h_{n-1}}(f)P-
P_{h_1,\dots,h_{n-1}}\wedge(gP_f+fP_g)=\nn\cr
&&P_{h_1,\dots,h_{n-1}}(fg)P-P_{h_1,\dots,h_{n-1}}\wedge((fg)\rfloor P)=
(fg)\rfloor(P_{h_1,\dots,h_{n-1}}\wedge P)
\eea
It remains to note that $P_{h_1,\dots,h_{n-1}}\wedge P=0$ for a multi-
vector of rank $n.\qquad\triangleright$

\begin{Cor}\label{C4}  Any Frobenius $n$-vector field $V$ on a manifold $M$ is an 
$n$-Pois\-son one. In particular, such is any $n$-vector field
on an $n$-dimensional manifold $M$.
\end{Cor}
{\it Proof.} Since $V$ defines an $n$-dimensional distribution
(with singularities) on $M$ it can be locally presented as
$V = hX_1\wedge...\wedge X_n$ for a suitable 
$h\in C^\infty(M)$. 
But $X_1\wedge...\wedge X_n$ is just the Poisson structure of 
example $\ref{E6}$ and, so, 
$V$ is also an $n$-Poisson structure in virtue of proposition $\ref{P6}$.$\triangleright$

\section{Decomposability of $n$-Poisson structures.}

  In this section we prove a result which, in a sense, is an analogue of
the Darboux lemma for $n$-Poison structures with $n>2$. It tells that the 
range of a non-trivial Poisson $n$-vector is equal to $n$ and, therefore, 
such a $n$-vector is locally decomposable. This was conjectured by Takhtajan 
and proved recently by Alexeevsky and Guha \cite{AG96}. Our approach is, however, 
quite different.
   We start with collecting  and recalling some elementary facts of 
multi-linear algebra. 
\par
Let ${\cal V}$ be a finite dimensional vector space. 
Denote by $\Lambda^k({\cal V})$ its $k$-th exterior power and put for 
$V\in \Lambda^k({\cal V})$ and $a_1,...,a_l\in {\cal V}^*$
\be
V_{a_1,...,a_l}:= a_l\rfloor~ ...\rfloor~ a_1\rfloor~ V\in \Lambda^k({\cal V})\nn
\ee  

The following is well-known.
  
\begin{Lem}\label{L1} 
A non-zero $k$-vector $V\in \Lambda^k({\cal V})$ is decomposable, 
i.e. $V = v_1\wedge...\wedge v_k,$  for some $v_i\in {\cal V}$, iff it is 
of rank $k$.
\end{Lem} 
Vectors  $v_i$'s are defined uniquely up to an unimodular transformation 
$v_i\ra w_i = \sum_j c_{ij}v_j$. The subspace of ${\cal V}$ generated by 
$v_1,...,v_k$  
coincides with that generated by all vectors of the form 
$V_{a_1,...,a_{k-1}}\in {\cal V}$ . 
   
Recall also the 
\begin{Lem}\label{LC} 
If 
$v\wedge V = 0, v\in {\cal V}, V\in \Lambda^k({\cal V})$, then $V$ 
is factorized by $v$, i.e. $V = v\wedge V'$ for a $V'\in \Lambda^k({\cal V})$ . 
\end{Lem}
Together with lemma $\ref{L1}$ this implies the following .

\begin{Lem}\label{L2} 
A $k$-vector $V$ is decomposable iff 

$V_{a_1,...,a_{k-1}}\wedge V = 0$,~~~ $\forall a_1,...,a_{k-1}\in {\cal V}^\ast$. 
\end{Lem} 
\begin{Lem}\label{L3} 
(on 3 planes). Let $\Pi_1,\Pi_2,\Pi_3$ be 
$(k-1)$-dimensional subspaces of 
${\cal V}$ such that $dim (\Pi_i\cap \Pi_j) = k-2$ for $i\neq j$. 

If $k>2$, then
\begin{itemize}
\item   
the span $\Pi$ of $\Pi_1,\Pi_2,\Pi_3$  is $k$-dimensional,
\item   
any $(k-1)$-dimensional subspace $\Pi'$ of ${\cal V}$ intersecting each of
 $\Pi_i$'s along not less than a $(k-2)$- dimensional subspace belongs to $\Pi$.
\end{itemize}
\end{Lem}  
{\it  Proof} Obvious. $\triangleright$

\begin{Prop}\label{P7} 
Let $V$ be a $k$-vector, $k>2$. If
\be\label{e14}
 V_{a,c_1,...,c_{k-2}}\wedge V_b + V_{b,c_1,...,c_{k-2}}\wedge V_a = 0,~~~
\forall a,b, c_1,...,c_{k-2}\in {\cal V}^*, 
\ee
then $V$ is decomposable.
\end{Prop} 
{\it Proof.} By putting $a = b$ in ($\ref{e14}$) we see that  
$W_{c_1,...,c_{k-2}}\wedge W = 0$ for $W = V_a$. 
Therefore, according to lemma $\ref{L2}$, the $(k-1)$-vector $V_a$ 
is decomposable $\forall a\in {\cal V}^*$.
\par   
Denote now by $\Pi_a$ the $(k-1)$-dimensional subspace of ${\cal V}$ 
canonically associated, according to lemma $\ref{L1}$, with the decomposable 
$(k-1)$-vector $V_a$ required to be different from zero.  
If $V_{a,c_1,...,c_{k-2}}\wedge V_b = 0$ for all 
$c_1,...,c_{k-2}\in{\cal V}\ast$, then $\Pi_a=\Pi_b$ as it results 
from lemma $\ref{LC}$ and lemma $\ref{L1}$.
If,otherwise, $V_{a,c_1,...,c_{k-2}}\wedge V_b \neq 0$ consider the subspace 
$\Pi$ associated according to lemma $\ref{L1}$ with the decomposable $k$-vector 
$V_{a,c_1,...,c_{k-2}}\wedge V_b $. Obviously, $\Pi\supset \Pi_b$. 

On the other hand, equality ($\ref{e14}$) shows that  $\Pi$ coincides 
with the sub\-space associ\-ated with the decompos\-able $k$-vector 
$V_{b,c_1,...,c_{k-2}}\wedge V_a = 0$. 
By this reason $\Pi\supset \Pi_a$ and, therefore, 
$dim(\Pi_a\cap \Pi_b)\geq k-2 > 0$. 
Moreover, if $V_{a,b}\neq 0$, then $dim(\Pi_a\cap \Pi_b) = k-2$. 
In fact, $dim(\Pi_a\cap \Pi_b) = k-1$ implies that $\Pi_a=\Pi_b$ 
and, hence, 
$V_a = \lambda V_b$ from which $V_{a,b} = \lambda V_{b,b}= 0$
what is impossible.
\par   
Observe, finally, that since $V\neq 0$ and $k\geq 3$ there exist 
$a,b,c\in {\cal V}^*$ such that $V_{a,b,c}\neq 0$. In such a situation 
$(k-2)$-vectors $V_{a,b}$, $V_{b,c}$ and $V_{a,c}$   are different 
from zero. Hence, as we have already previously seen, mutual intersections
$\Pi_a$, $\Pi_b$ and $\Pi_c$ are all $(k-2)$-dimensional. 
So, these three subspaces satisfy the hypothesis of lemma $\ref{L3}$. 
By this reason the span $\Pi$ of them contains all subspaces 
$\Pi_d$,~~ $d\in {\cal V}^\ast$, and consequently all 
derived vectors $V_{d,d_1,...,d_{k-2}}$ belong to $\Pi$. 
Now lemma $\ref{L1}$ implies the desired result. $\triangleright$
\par   
Our next task is to show that the hypothesis of proposition $\ref{P7}$ is 
satisfied by any Poisson multi-vector.
   First, we need the following property of Lie derivations.
\begin{Lem}\label{L4} 
Let $X\in {\cal D}(M)$ and $f\in C^\infty(M)$. 
For a multi-derivation $\Delta$ it holds
\be\nn
L_{fX}(\Delta) = fL_X(\Delta)-X\wedge \Delta_f
\ee
\end{Lem}  
{\it  Proof.} By the definition of the Lie derivative we have
\bea\nn
L_{fX}(\Delta)(g_1,...,g_n)&=&fX(\Delta(g_1,...,g_n))
-\sum_i\Delta(g_1,...,fX(g_i),...,g_n)\cr 
&=&f(X(\Delta(g_1,...,g_n)-\sum_i\Delta(g_1,...,X(g_i),...g_n))\cr 
&-&\sum_i(-1)^{i-1} X(g_i)\Delta(f,g_1,...,g_n).
\eea
It remains to note that last sum is just the product $X\wedge \Delta_f$
 evaluated on $g_1,...,g_n$. $\triangleright$
   
Next identity is basic.
   
\begin{Prop}\label{P8} 
Let $\Delta$ be an $n$-derivation. 
Then for any $f,g,\phi_i\in C^\infty(M)$ 
it holds:
\bea\label{e15}
\Delta_{fg,\phi_1,...,\phi_{n-2}}(\Box)=f\Delta_{g,\phi_1,...,\phi_{n-2}}(\Box)+\nn\\
g\Delta_{f,\phi_1,...,\phi_{n-2}}(\Box)-\Delta_{f,\phi_1,...,\phi_{n-2}}\wedge\Box_g-\nn\\
\Delta_{g,\phi_1,...,\phi_{n-2}}\wedge\Box_f 
\eea   
\end{Prop}
{\it Proof.} 
First, note that $\Delta_{fg} = f\Delta_g + g\Delta_f$  so that one has
\bea\nn
\Delta_{fg,\phi_1,..,\phi_{n-2}}(\Box)&=&  
(f\Delta_{g,\phi_1,..,\phi_{n-2}})(\Box) +
(g\Delta_{f,\phi_1,..,\phi_{n-2}})(\Box)
\eea
On the other hand, by putting $Y = \Delta_{f,\phi_1,...,\phi_{n-2}}$, 
$Z = \Delta_{g,\phi_1,...,\phi_{n-2}}$ 
and applying lemma $\ref{L4}$ one finds
\bea\label{e16}
\Delta_{fg,\phi_1,...,\phi_{n-2}} (\Delta_f)= (L_{gY} + L_{fZ})(\Box)=\cr
gL_Y(\Box) +fL_Z(\Box) - Y\wedge \Box_g - Z\wedge \Box_f \qquad\triangleright 
\eea
\begin{Cor}\label{C5} 
If $\Delta$ is an $n$-Poisson structure, then for any 
$f,g,\phi_i\in C^\infty(M)$ 
it holds
\be\label{e18}      
\Delta_{f,\phi_1,...,\phi_{n-2}}\wedge 
\Delta_g + \Delta_{g,\phi_1,...,\phi_{n-2}}\wedge \Delta_f = 0          
\ee
\end{Cor}   
{\it Proof.}  
Formula ($\ref{e15}$) for an $n$-Poisson  $\Delta$, and $\Box=\Delta$ 
is reduced, obviously, to (\ref{e18})$\qquad\triangleright$

\begin{emrem}\label{R5} 
Formula ($\ref{e18}$) for $n = 2$ becomes empty.
   We mention also the following particular case of ($\ref{e18}$) for which $g = f$ :
\be\label{e19}
 \Delta_{g,\phi_1,...,\phi_{n-2}}\wedge \Delta_g = 0                       
\ee
\end{emrem}   
\begin{Th}\label{Th1}
 Any non trivial $n$-Poisson $n$-vector $V$ 
is of rank $n$ if $n > 2$.
\end{Th}  
{\it  Proof.} Formula ($\ref{e18}$) can be rewritten as
\bea\nn
&(d\phi_{n-1}\rfloor~...d\phi_1\rfloor~df\rfloor~V)\wedge 
(dg\rfloor~V) +\cr &(d\phi_{n-1}\rfloor~...d\phi_1\rfloor~dg\rfloor~V)
\wedge (df\rfloor~V) = 0
\eea
Evaluated at a point $x\in M$ it ensures the hypothesis ($\ref{e14}$) 
of proposition $\ref{P7}$ for the $n$-vector $V_x$ over the tangent space
$ {\cal V} = T_xM$. 
Therefore, $V_x$ is of rank $n$ or identically equal to zero, 
otherwise.$\quad\triangleright$
   
\begin{Cor}\label{C6} For $n>2$ regular leaves of the Hamiltonian foliation of 
an $n$-Poisson manifold are $n$-dimensional. Its singular 
leaves are just points. 
\end{Cor}
\begin{emrem}\label{R6} Since an $n$-dimensional foliation can be given by means of
 $n$ commuting vector fields in a neighborhood of its regular point, 
example $\ref{E6}$ exhausts regular local forms of $n$-Poisson structures for $n > 2$.
\end{emrem}   
Another eventually very important consequence of theorem $\ref{Th1}$ is that 
the cartesian product of two $n$-Poisson manifolds is not in a natural 
way a such one if $n > 2$. In fact, there is no natural way to construct 
an $n$-dimensional foliation on the cartesian product of two manifolds 
supplied with such ones.

Theorem $\ref{Th1}$ shows $n$-Poisson structures for $n>2$ to be extremely
rigid what implies some peculiarities going beyond the {\it binary}
based expectations. Below we exhibit two of them: no cartesian
products and no (in general) $n$-Poisson structure on the dual 
 of an $n$-Lie algebra.
\par
First, note that given two $n$-vector fields $P$ and $Q$ on manifolds 
$M$ and $N$, respectively, their direct sum $P\oplus Q$ which is
an $n$-vector field on $M\times N$ is naturally defined.

\begin{Cor}\label{C6+1} If $P$ and $Q$ are non-trivial $n$-Poisson vector
 fields, then $P\oplus Q$ is not an $n$-Poisson one for $n>2$.
\end{Cor}
{\it Proof.} Just to note that rank $(P\oplus Q)=\mbox{rank}(P)+
\mbox{rank}(Q).\qquad\triangleright$
\par
This result can be also proved by a direct computation.
\par
Second, given an $n$-Lie group structure $[\cdot,\dots,\cdot]$
on ${\cal V}$ one can try to associate with it an $n$-Poisson
structure on its dual ${\cal V}^*$ just by copying the standard construction
for $n=2$. Namely, let $x_1,\dots,x_N\in {\cal V}$ be a basis. 
Interpreting $x_i$'s to be coordinate functions on ${\cal V}^*$, let us put
\be\label{Poo}
T=\sum_{1\leq i_{1}<\dots<i_{n}\leq N}[x_{i_1},\dots,x_{i_n}]
\frac{\partial}{\partial x_{i_1}}\wedge\dots\wedge
\frac{\partial}{\partial x_{i_n}}
\ee

In a coordinate-free form the $n$-vector field $T$ 
can be presented as
$$
T(df_1,\dots,df_n)(u)=[d_{u}f_1,\dots,d_{u}f_n]
$$
with $u\in {\cal V}^*$ and $f_i\in {\cal C}^\infty({\cal V}^*)$ 
where the differential
$d_{u}f_i$   of $f_i$ at the point $u$ is interpreted 
canonically to be an element of ${\cal V}$. This $n$-vector field  
$T$ is called {\it associated} with the $n$-Lie algebra
structure in question.  

 It is well known (for instance, \cite{Vin-Kra}) this 
formula defines the standard Poisson structure on ${\cal V}^*$
when $n=2$. However, it is no longer so when $n>2$.

\begin{Cor}\label{C6+2} If $n>2$ the $n$-vector field $T$ given
 by ($\ref{Poo}$) is not generally an $n$-Poisson one.
\end{Cor}
{\it Proof.} First, note that the $n$-vector field associated
with the direct product of two $n$-Lie algebras is the direct sum
of $n$-vector fields associated with each of them.
Since, obviously, the $n$-vector field associated with a non-trivial 
$n$-Lie algebra is of rank not less than $n$, the $n$-vector field 
associated with the product of two non-trivial $n$-Lie algebras is of 
rank not less $2n$. Therefore, it cannot be an $n$-Poisson vector 
if $n>2 \qquad\triangleright$.
\par
On the other hand we have:
 
\begin{Prop}\label{P8+1}
Formula ($\ref{Poo}$) defines an $n$-Poisson 
structure on the dual of an  $n$-Lie algebra of dimension 
$\leq n+1$.
\end{Prop}
{\it Proof.}  As it easy to see any $n$-vector defined on a 
space of dimension $\leq n+1$ is  either of rank $n$ or 0. 
So, under the hypothesis of the proposition $T$ defines an 
$n$-or 0-dimensional distribution on ${\cal V}^*$. 
Denote by   $\Delta$ the $n$-derivation on
${\cal V}^*$ corresponding to $T$ as in ($\ref{Poo}$). 
It suffices to show that
\be\label{AXO}
\Delta_{f_1,\dots,f_{n-1}}(\Delta)=0
\ee
for any system of polynomials $f_i(x)$ in variables $x_k$'s.
We prove it by induction on the total degree
$\delta= degf_1+\dots+ degf_{n-1}$ by starting
from $\delta=n-1$.

To start the induction note that in the case
all $f_i$'s are linear on ${\cal V}^*$, i.e. elements of ${\cal V}$,
identity (\ref{AXO}) is identical to the $n$-Jacobi identity 
of the original $n$-Lie algebra.

To complete the induction it is sufficient to show that
(\ref{AXO})  holds for the system $f_1=gh,f_2,\dots,f_{n-1}$
if it holds for $g,f_2,\dots,f_{n-1}$  and  $h,f_2,\dots,f_{n-1}$.
Taking into account that 
$\Delta_{gh,f_2,\dots,f_{n-1}}=g\Delta_{h,f_2,\dots,f_{n-1}}
+h\Delta_{g,f_2,\dots,f_{n-1}}$ and lemma $\ref{L4}$ the problem 
is reduced to prove that
\be\label{AXY}
\Delta_{g,f_2,\dots,f_{n-1}}\wedge\Delta_h+\Delta_{h,f_2,\dots,f_{n-1}}
\wedge\Delta_g=0
\ee
But since $T$ is of rank $n$ $\Delta_{\varphi_1,\dots,\varphi_{n-1}}
\wedge \Delta=0$ for any system $\varphi_1,\dots,\varphi_{n-1}\in {\cal C}^\infty({\cal V}^*)$
we have 
$$
0=h\mathbin{\rfloor}(\Delta_{g,f_2,\dots,f_{n-1}}\wedge\Delta)=
\Delta(g,f_2,\dots,f_{n-1},h)\Delta-\Delta_{g,f_2,\dots,f_{n-1}}
\wedge\Delta_h,
$$
so that

$\Delta_{g,f_2,\dots,f_{n-1}}\wedge\Delta_h=
\Delta(g,f_2,\dots,f_{n-1},h),
$ 

and, similarly,

$
\Delta_{h,f_2,\dots,f_{n-1}}\wedge\Delta_g=\Delta(h,f_2,\dots,f_{n-1}g).
$

Hence, ($\ref{AXY}$) results from skew sym\-metry of $\Delta.\qquad\triangleright$

Previous discussions leads us to conjecture that:
\par
{\it If $n>2$ any $n$-Lie algebra is split into the direct product of
a trivial  $n$-Lie algebra and a number of non-trivial $n$-Lie algebras
of dimensions $n$ and 
$(n+1)$}.
\par
In fact, this conjecture saves in essence the fact that an $n$-Lie
algebra structure generates an $n$-Poisson structure on its dual 
(Proposition $\ref{P8+1}$) in view of the 
resistance of $n$-Poisson manifolds to form cartesian products
(corollary $\ref{C6+1}$) if $n>2$.  Also, at least to our knowledge,
all known examples in the literature  are in favor of this conjecture.

Finally, mention an alternative (and also natural) way to save
the dual construction by giving to the concept of $n$-Poisson
manifold the {\it dual} meaning (see \cite{VV}). For fundamentals 
of this dual approach we send the reader to \cite{Mich Vin}.
A discussion of the Koszul duality can be found in \cite{Loday}
and \cite{Gned}. 

We conclude this section by answering the natural question:
what are {\it multiplicative} compatibility conditions for two 
multi-Poisson structures, i.e. conditions ensuring that 
their wedge product is again a multi-Poisson one.

\begin{Prop}\label{P8+2} 
Let $\Delta$ and $\nabla$ be the multi-
Poisson structures on the manifold $M$ whose multiplicities coincide
with their rank (for instance, they are of multiplicities greater
than two.) Then $\Delta\wedge\nabla$ is a multi-Poisson structure on
$M$ iff $\lceil\Delta,\nabla\rfloor=0,\quad
\Delta_{g_1,\dots,g_{k-1}}(\nabla)\wedge\nabla=0,\quad\nabla_{h_1,
\dots,h_{l-1}}(\Delta)\wedge\Delta=0$ for all $g_i,h_j\in{\cal C}^\infty(M)$,
$k$ and $l$ denoting the multiplicities of $\Delta$ and $\nabla$,
respectively.
\end{Prop}
{\it Proof.} First, note the formula which is a direct consequence of 
the wedge product definition:
\be\label{IWP}
(\Delta\wedge\nabla)_{f_1,\dots,f_N}=
\sum_{I}(-1)^{(k-|I|)(N-|I|)+(I,{\bar I})}\Delta_{f_I}\wedge\nabla_{f_{\bar I}}
\ee
where $I$ runs all ordered subsets of $\{1,\dots,N\}$ and $|I|$ denotes 
the cardinality of $I$.
In particular, for $N=k+l-1$  we have
\be\label{PWP}
(\Delta\wedge\nabla)_{f_1,\dots,f_{k+l-1}}=
\sum_{|I|=k}(-1)^{(I,{\bar I})}\Delta(f_I)\nabla_{f_{\bar I}}+
\sum_{|I|=k-1}(-1)^{l+(I,{\bar I})}\nabla(f_{\bar I})\Delta_{f_I} 
\ee
By applying lemma $\ref{L4}$ to $f=\Delta(f_I),X=\nabla_{f_{\bar I}}$ and
taking into account that $\nabla_{f_{\bar I}}(\nabla)=0$ and 
$\nabla_{f_{\bar I}}\wedge\nabla=0\quad(\nabla$ is $l$-Poisson
of rank $l$) we find
\be\label{ING}
(\Delta(f_I)\nabla_{f_{\bar I}})(\Delta\wedge\nabla)=
\Delta(f_I)\nabla_{f_{\bar I}}(\Delta)\wedge\nabla-(-1)^k\nabla_{f_{\bar I}}
\wedge\Delta\wedge\nabla_{\Delta(f_I)}
\ee
and, similarly,
\be\label{GNI}
(\nabla(f_{\bar I})\Delta_{f_I})(\Delta\wedge\nabla)=
\nabla(f_{\bar I})\Delta\wedge\Delta_{f_I}(\nabla)-
\Delta_{f_I}\wedge\Delta_{\nabla(f_{\bar I})}\wedge\nabla
\ee
Since $\Delta_{f_I}\wedge\Delta=0$, then
$$
0=\nabla(f_{\bar I})\rfloor(\Delta_{f_I}\wedge\Delta)=
\Delta(f_I,\nabla(f_{\bar I}))\Delta-
\Delta_{f_I}\wedge\Delta_{\nabla(f_{\bar I})}
$$
that is
\be\label{IJO}
\Delta_{f_I}\wedge\Delta_{\nabla(f_{\bar I})}=\Delta(f_I,\nabla(f_{\bar I}))
\ee
and, similarly,
\be\label{JIO}
\nabla_{f_{\bar I}}\wedge\nabla_{\Delta(f_I)}=\nabla(f_{\bar I},\Delta(f_I))
\ee
Now bearing in mind (\ref{PWP})-(\ref{JIO}) we get
\bea\label{YX0}
&&(\Delta\wedge\nabla)_{{f_1},\dots,f_{k+l-1}}(\Delta\wedge\nabla)\cr
&&=\sum_{|I|=k}(-1)^{(I,{\bar I})}(\Delta(f_I)\nabla_{f_{\bar I}}(\Delta)
\wedge\nabla-\nabla(f_{\bar I},\Delta(f_I))\Delta\wedge\nabla)\cr
&&+\sum_{|I|=k-1}(-1)^{l+(I,{\bar I})}(\nabla(f_{\bar I})\Delta\wedge\Delta_{f_{I}}
(\nabla)-\Delta(f_I,\nabla(f_{\bar I}))\Delta\wedge\nabla)\cr
&&=\sum_{|I|=k}(-1)^{(I,{\bar I})}\Delta(f_I)\nabla_{f_{\bar I}}(\Delta)\wedge\nabla
+\sum_{|I|=k-1}(-1)^{l+(I,{\bar I})}\nabla(f_{\bar I})\Delta\wedge\Delta_{f_I}
(\nabla)\cr
&&-(-1)^{l}\lceil \Delta,\nabla\rfloor(f_1,\dots,f_{k+l-1})
\eea
(see ($\ref{e12}$)). If $\Delta\wedge\nabla$ is a multi-Poisson structure,
then it is also a multi-Poisson structure in the dual sense defined in 
 \cite{Mich Vin}.
But for such structures $\Delta\wedge\nabla$ is multi-Poisson iff
$\lceil \Delta,\nabla\rfloor=0$. This shows that   $\lceil \Delta,\nabla\rfloor=0$
is a necessary condition for the considered problem.
\par
Observe now that due to local decomposability of multi-vector
corresponding to $\Delta$ and $\nabla$ the product $\Delta\wedge\nabla$ 
is different from zero iff they are transversal to each other. 
This implies that the leaves of the
corresponding Hamiltonian foliations intersect one another transversally.
By this reason one can find $k$ local Casimir functions of $\nabla$, say
$f_1,\dots,f_k$, such that $\Delta(f_1,\dots,f_k)\neq 0$ and $l$ Casimir
functions of $\Delta$, say $f_{k+1} ,\dots,f_{k+l}$  such that 
$\nabla(f_{k+1},\dots,f_{k+l})\neq 0$.
For such chosen $f_i$'s all summands of the first two summations of
($\ref{YX0}$) vanish except one which is 
$$\Delta(f_1,\dots,f_k)\nabla_{f_{k+1},\dots,f_{k+l-1}}(\Delta)\wedge\nabla.$$
This implies $\nabla_{f_{k+1},\dots,f_{k+l-1}}(\Delta)\wedge\nabla=0$,~ 
if $\Delta\wedge\nabla$ is $(k+l)$-Poisson. Observing then that local 
Casimir functions of both $\Delta$ and $\nabla$ generate in that 
situation a local smooth function algebra, one can conclude that
\be\label{N11}
\nabla_{g_1,\dots,g_{l-1}}(\Delta)\wedge\nabla=0
\ee
for any family of functions $g_1,\dots,g_{l-1}$.

Similarly, it is proved that
\be\label{N22}
\Delta_{g_1,\dots,g_{l-1}}(\nabla)\wedge\Delta=0
\ee
This shows that (\ref{N11}),(\ref{N22}) and $\lceil \Delta,\nabla\rfloor=0$
 are necessary. Their sufficiency is obvious from ($\ref{YX0}$).$\qquad\triangleright$

\section{Local $n$-Lie algebras.}

In this section we discuss the most general natural synthesis of the 
concept of multi-Lie algebra  and that of smooth manifold which is as 
follows.
\par
\begin{Def}\label{D5} A local $n$-Lie algebra structure on a manifold M is an 
$n$-Lie algebra structure 
$$
(f_1,...,f_n)\ra  [f_1,...,f_n]
$$
on $C^\infty(M)$ which is a multi-differential operator. 
\end{Def}
Below we continue to use the {\it operator} notation as well as the 
{\it bracket} one for local $n$-ary structures :
$$
\Delta(f_1,...,f_n) = [f_1,...,f_n]
$$
and refer to the multi-differential operator $\Delta$ as the structure 
in question itself.
  
\begin{emexa}\label{E8} 
$n$-Poisson structures are local $n$-Lie algebra ones.
\end{emexa} 
A well-known result by Kirillov \cite{Ki76} says that for $n = 2$ the 
bi-differential operator giving a local Lie algebra structure on a 
manifold $M$ is of first order with respect to both its arguments. 
An interesting algebraic proof of this fact can be found in \cite{GMP93}. 
Kirillov' s theorem is generalized immediately to higher local 
multi-Lie algebras.
  
\begin{Prop}\label{P9} 
Any local $n$-Lie algebra, $n\geq 2$ is given by an 
$n$-differential operator of first order, i.e. of first order with 
respect each its argument.
\end{Prop}   
{\it Proof.} It results from Kirillov's theorem applied to $(n-2)$-order
hereditary structures of the considered algebra.$\qquad\triangleright$ 
   
Recall that usual Lie algebra structures defined by means of 
first order bi-differential operators are called {\it Jacobi}'s 
\cite{Ki76,Ly,KLV}. This motivates the following terminology.
   
\begin{Def}\label{D6} An $n$-Jacobi manifold (structure) is a manifold M 
supplied with a local $n$-Lie algebra structure on $C^\infty(M)$ given by a 
first order $n$-differential operator.
\end{Def}
Hence, in these terms proposition $\ref{P9}$ says that multi-Jacobi structures
exhaust local multi-Lie algebra ones. Note, however, that it seems not 
to be the case for infinite dimensional manifolds such that occur in 
Secondary Calculus. Kirillov gives also an exhaustive description  
of Jacobi manifolds. 
\par
Namely, Kirillov showed that a binary Jacobi bracket $[\cdot~,~\cdot]$
 on a manifold $M$ can be uniquely presented in  the form
$$
[f,g]=T(df,dg)+fX(g)-gX(f)
$$
with $X$ and $T$ being a vector field and a bivector field,
respectively, such that 
$\|T,T\|=X\wedge T$ and $L_X(T)=0$
Then two qualitative different situations can occur:
$X\wedge T\equiv 0$ and $X\wedge T\ne 0$ (locally). 
In the first of them the bivector $T$ is a Poissonian of rank 0 or 2.
 In the latter case $X$ is a locally Hamiltonian field with  
respect to $T$, i.e. $X=T_f$ for an appropriate $f\in{\cal C}^\infty (M)$.
If $X\wedge T\ne 0,$then $M$ is foliated (with singularities)
by ($2n+1$)-dimensional leaves with $2n=\mbox{rank} T$ 
(an analog of the Hamiltonian foliation) and the original Jacobi structure
is reduced to a family of locally contact brackets \cite{Ly,KLV} on leaves
of this foliation.
\par
Below we find an $n$-ary analogue of Kirillov's theorem for $n>2$
showing that in this case only the first possibility of two
mentioned above survives.
Fundamental here is a canonical decomposition of the first order 
skew-symmetric multi-differential operator $\Delta$ defining the local 
$n$-Lie algebra in question which we are passing to describe.
   
Recall, first, that a first order linear (scalar) differential 
operator on $M$ is a ${\bf R}$-linear map  
$\nabla: C^\infty (M)\ra  C^\infty (M)$ 
such that
\be
\nabla(fg) =f\nabla (g)+g\nabla (f)-fg\nabla (1)~~~~\forall f,g\in C^\infty (M).
\ee
This al\-gebraic definition is equiva\-lent to the standard coordinate
one \cite{KLV}. It  characterizes vector fields on $M$, i.e. 
derivations of $C^\infty (M)$, as first order differential 
operators $\nabla$ such that $\nabla(1) = 0$.
   Let $\Delta$ be a skew-symmetric first order $n$-differential operator. 
According to the adopted notation $\Delta_1$ is an $(n-1)$-differential 
operator defined as $\Delta_1(f_1,...,f_{n-1}) = \Delta(1,f_1,...,f_{n-1})$. 
Obviously, it is of first order. Moreover, it is a multi-derivation. 
In fact, it is seen immediately from what was said before by observing 
that owing to skew-commutativity 
\be\nn
\Delta_1(1,...) = (\Delta_1)_1 = \Delta_{1,1} = 0
\ee
  If $\Gamma$ is a skew-symmetric $k$-derivation, then the $(k+1)$-differential 
operator $s(\Gamma)$ defined as
\be\nn
s(\Gamma)(f_1,...,f_{k+1}) = 
\sum_i(-1)^{i-1}f_i\Gamma(f_1,...,f_{i-1},f_{i+1},...,f_{k+1}).
\ee
is, obviously, skew-symmetric and of first order. 
Moreover, $s(\Gamma)_1 = \Gamma$.
   By applying this construction to $\Gamma = \Delta_1$ we obtain the first 
order skew- symmetric $n$-differential operator $\Delta^0=s(\Delta^1)$ such that 
($\Delta^0)_1 = \Delta_1$. 
Last relation shows that the $n$-differential operator 
$\hat\Delta= \Delta - \Delta^0$, is 
an $n$-derivation. Now gathering together what was done before we obtain :

\begin{Prop}\label{P10} 
With any first order skew-symmetric 
$n$-differential operator
 $\Delta$ are associated skew-symmetric multi-derivations 
$\Delta$ and $\Delta_1$
 of multiplicities $n$ and $n-1$, respectively, such that 
({\it canonical decomposition})
\be\nn 
\Delta= \hat\Delta + \Delta^0        
\ee
with $\Delta^0= s(\Delta_1)$, i.e.
\be\nn
\Delta^0 (f_1,...,f_n) = \sum_i(-1)^{i-1}f_i
\Delta_1(f_1,...,f_{i-1},f_{i+1},...,f_n).
\ee
Conversely, any pair $(\nabla,\Gamma)$ of skew-symmetric derivations of 
multiplicities $n$ and $n-1$, respectively, defines an unique skew-symmetric
$n$-differential operator of first order  $\Delta = \nabla+s(\Gamma)$
such that $\nabla = \hat\Delta$  and $\Gamma = \Delta_1$. $\triangleright$
\end{Prop}
It is natural to extend the operation $s$ from the skew-symmetric derivations
to arbitrary skew-symmetric multi-differential operators. Namely, if $\Delta$
is a skew-symmetric $k$-differential operator, then we put
\bea\nn
s(\Delta)(g_1,...,g_{k+1}) = 
\sum\limits_{i=1}^{k+1} (-1)^{i-1} g_i\Delta (g_1,...,g_{i-1},
g_{i+1},...,g_{k+1})
\eea
This way we get the map
\bea\nn
s:{ Diff}_{l|k}^{alt} (M) \ra {Diff}_{l|k+1}^{alt} (M), 
\eea
${Diff}_{l|k}^{alt} (M)$ denoting the space of~$l$-th order 
${\cal C}^{\infty} (M)$
-valued skew-sym\-metric $k$-differential operators on  ${\cal C}^{\infty} (M)$.

\begin{Prop}\label{P11} 
The operation $s$ is ${\cal C}^{\infty} (M)-$linear and $s^2=0$.
\end{Prop}
{\it Proof.} Obvious. $\triangleright$

\begin{emrem}\label{R7} 
Proposition ($\ref{P11}$) shows that $s$ can be viewed as the differential 
of the complex:
\bea\nn
0 \ra {Diff}_{l|1}^{alt} (M)\stackrel{s}{\ra} 
{Diff}_{l|2}^{alt} (M)\stackrel{s}{\ra}...\stackrel{s}{\ra}
{Diff}_{l|k}^{alt} (M)\stackrel{s}{\ra}...
\eea
This complex is acyclic in positive dimensions and its 0-cohomology group 
is isomorphic to
 ${\cal C}^{\infty} (M)$. 
In fact, the insertion of the unity operator  $i_1$ is a homotopy operator 
for $s$ as it
results from Proposition $\ref{P10}$.
\end{emrem}
Further properties of $s$ we need are the following.

\begin{Prop}\label{P12} 
The operation $s$ has the properties:
\begin{enumerate}
\item If $\;X\in D(M),$ then $[L_{X},s]=0$
\item If $\;f\in{\cal C}^{\infty} (M),\;$ then $\;f\rfloor s(\Box)+s (f\rfloor \Box) = 
f \Box\;$ and
\item $\;s(\Box)_{f_1,...,f_k} = \sum_i (-1)^{i-1}f_i\Box_{f_1,...,f_{i-1},f_{i+1},...,f_k}+
(-1)^k\Box(f_1,...,f_k)$
\end{enumerate}
\end{Prop}
{\it Proof.} We start with number 1. 

For $\;\Box \in{Diff}_{l|k}^{alt} (M)\;$
one has by definition
\bea\nonumber
 L_X(s(\Box))(g_1,..,g_{k+1})&=&X(s(\Box)(g_1,..,g_{k+1}))\cr 
&-&\sum_{i} s(\Box)(g_1,..,{X}(g_i),..,g_{k+1})
\eea 
But
\bea
X(s(\Box)(g_1,..,g_{k+1}))=
\sum_i(-1)^{i-1} X(g_i)\Box(g_1,..,g_{i-1},g_{i+1},..,g_{k+1})\nonumber\\
+\sum_i (-1)^{i-1}g_i X(\Box(g_1,..,g_{i-1},g_{i+1},..,g_{k+1})\nonumber
\eea
and
\bea
s(\Box)(g_1,..,X(g_i),...,g_{k+1})= 
(-1)^{i-1} X(g_i)\Box(g_1,..,g_{i-1},g_{i+1},..,g_{k+1})\nonumber\\
+\sum_{j<i}(-1)^{j-1}g_j\Box(g_1,..,g_{j-1},g_{j+1},..,X(g_i),..,g_{k+1})\nonumber\\+
\sum_{i<j}(-1)^{j-1}g_j\Box(g_1,..,X(g_i),..,g_{j-1},g_{j+1},..,g_{k+1})\nonumber
\eea

Therefore,
\bea
 L_X(s(\Box))(g_1,...,g_{k+1})=
\sum_i(-1)^{i-1}g_i X(\Box(g_1,...,g_{i-1},g_{i+1},...,g_{k+1})\nonumber\\
+\sum_{j<i}(-1)^{j-1}g_j\Box(g_1,...,g_{j-1},g_{j+1},...,X(g_i),...,g_{k+1})\nonumber\\
+\sum_{i<j}(-1)^{j-1}g_j\Box(g_1,...,X(g_i),...,g_{j-1},g_{j+1},...,g_{k+1})\nonumber\\
=\sum_i(-1)^{i-1}g_i X(\Box)g_1,...,g_{i-1},g_{i+1},...,g_{k+1})=
s(L_X(\Box))(g_1,...,g_{k+1})\nonumber
\eea
Thus, $\;s\circ L_X = L_X\circ s \Leftrightarrow [ L_X, s] = 0$.

Property 2 is an immediate consequence of the definition of $\;s\;$.
Finally, 3 is obtained from 2 by an obvious induction. $\triangleright$

We need also the following formula concerning Lie derivative

\begin{Lem}\label{L5} If $\;f\in{\cal C}^{\infty} (M)\;$ and $\;\Box\;$ 
is a skew-symmetric $k$-derivation, then
\bea\nonumber
 L_f(\Box)=(1-k)f\Box-s(\Box_f)
\eea
where the {\it Lie derivative} $ L_f$ is understood in the sense of section 2.
\end{Lem}
{\it Proof.}
By definition
\bea
&&L_f(\Box)(g_1,...,g_k)=
f\cdot\Box((g_1,...,g_k)-\sum(g_1,...,fg_i,...,g_k)\nonumber\cr
&&=f\cdot\Box(g_1,...,g_k)-
\sum_i(f\cdot\Box(g_1,...,g_k)+g_i\Box(g_1,...,g_{i-1},f,g_{i+1},...,g_k)\nonumber\cr
&&=(1-k)f\Box(g_1,...,g_k) + \sum(-1)^{i-1}g_i\Box(f,g_1,...,g_{i-1},g_{i+1}...,g_k)\nonumber
\eea
But last summation coincides, obviously, with  $-s(\Box_f)(g_1,...,g_k)$.$\quad\triangleright$
\par
Proposition $\ref{P10}$ suggests to treat the problem of describing $n$-Jacobi
structures as determination of conditions to impose on a pair of multi-derivations
$\nabla$ and $\Box$ of multiplicities $n$ and $n-1$, respectively, in order
the $n$-differential operator $\Delta =\nabla+s(\Box)$ be an $n$-Jacobi one.
In other words, we have to resolve the equation:
\be\label{ADD}
(\nabla+s(\Box))_{f_1,\dots ,f_{n-1}}(\nabla+s(\Box))=0
\ee
with respect to $\nabla$ and $\Box$. So we pass to analyze equation ($\ref{ADD}$)
\par
First, by applying proposition 
($\ref{P12}$, 3)
and posing $\; X_i=\Box_{f_1,...,{\hat f}_i,...,f_{n-1}}\;$ and 
$\;h=(-1)^{n-1}\Box(f_1,...,f_{n-1})\;$ one finds
$$
s(\Box)_{f_1,...,f_{n-1}}(\nabla)=\sum(-1)^{i-1}(f_i X_i)(\nabla)+ L_h(\nabla)
$$
The following expression is computed with the help of lemmas $\ref{L4}$ and $\ref{L5}$:
\be\label{A21}
s(\Box)_{f_1},...,f_{n-1}(\nabla)=
\sum(-1)^{i-1}(f_i X_i(\nabla)-X_i\wedge
\nabla_{f_i})+(1-n)h\nabla-s(\nabla_h)
\ee
Similarly, taking into account proposition ($\ref{P12}$,1), lemma $\ref{L4}$, 
lemma $\ref{L5}$ and the fact that 
$s(\Box)_{f_1,...,f_{n-1}}= Y+h$ with 
$Y=\sum(-1)^{i-1}f_i X_i\in{\cal D}(M)$
one finds 
\bea\label{A22}
s(\Box)_{f_1,...,f_{n-1}}(s(\Box))=
( Y+h)(s(\Box))=s(Y(\Box))+(1-n)hs(\Box)\nonumber\\
=s(\sum(-1)^{i-1}f_i X_i(\Box)-
\sum(-1)^{i-1} X_i\wedge\Box_{f_i}+(1-n)h\Box)
\eea
Putting together formulae ($\ref{A21}$) and ($\ref{A22}$) we obtain 
the key technical result of this section.

\begin{Prop}\label{P13} 
Let $\;\nabla\;$ and $\;\Box\;$ be skew-symmetric multi-derivations
of multiplicity  $n$ and $n-1$, respectively,
then the canonical decomposition of the skew-symmetric 
$k$-differential operator 
$$
(\nabla+s(\Box))_{f_1,...,f_{n-1}}(\nabla+s(\Box))$$
 is given by the formula: 
$$
(\nabla+s(\Box))_{f_1,...,f_{n-1}}(\nabla+s(\Box))=\Delta^1+s(\Delta^{o})
$$
where 
\bea
\Delta^1(f_1,...,f_{n-1})&=&\nabla_{f_1,...,f_{n-1}}(\nabla)\nonumber\cr
&+&\sum_{i=1}^{n-1}(-1)^{i-1}(f_i X_i(\nabla)-
 X_i\wedge\nabla_{f_i}+(1-n)h\nabla)\nonumber
\eea
and
\bea
\Delta^{o}(f_1,...,f_{n-1})&=&\nabla_{f_1,...,f_{k-1}}(\Box)-\nabla_h\nonumber\cr
&+&\sum_{i=1}^{n-1}(-1)^{i-1}(f_i X_i(\Box)- X_i\wedge\Box_{f_i}+(1-n)h\Box)\nonumber
\eea
with~~ $ X_i=\Box_{f_1,...,f_{i-1},f_{i+1},...,f_{n-1}},
\quad h=(-1)^{n-1}\Box(f_1,...,f_{n-1}).\qquad\triangleright$
\end{Prop}

\begin{Cor}\label{C7} If $\Delta+s(\Box)$ is $n$-Jacobian, then  
for any $g_1,...g_{n-2}\in{\cal C}^\infty(M)$
$$
\Box_{g_1,...,g_{n-2}}(\nabla)=0 \quad\mbox{and}\quad
\Box_{g_1,...,g_{n-2}}(\Box)=0
$$
In particular, $\Box$ is an ($n-1$)-Poisson structure.
\end{Cor}
{\it Proof.} In virtue of proposition $\ref{P13}$ equation ($\ref{ADD}$) 
is equivalent to
$$
\Delta^{o}(f_1,...,f_{n-1})=0\quad,\quad\Delta^1(f_1,...,f_{n-1})=0
$$
It remains to note that 
\bea
\Delta^{o}(1,g_1,...,g_{n-2})&=&\Box_{g_1,...,g_{n-2}}(\Box)\nonumber\cr
\Delta^{1}(1,g_1,...,g_{n-2})&=&\Box_{g_1,...,g_{n-2}}(\nabla)\nonumber\quad\triangleright
\eea

Put
\bea
\Delta_{o}^1(f_1,...,f_{n-1})&:=&\nabla_{f_1,...,f_{n-1}}(\nabla)+
\sum_{i=1}^{n-1}(-1)^{i-1}f_i\rfloor(X_i\wedge\nabla)\nonumber\cr
\Delta_{o}^o(f_1,...,f_{n-1})&:=&\nabla_{f_1,...,f_{n-1}}(\Box)+
\sum_{i=1}^{n-1}(-1)^{i-1}f_i\rfloor( X_i\wedge\Box)-\nabla_h\nonumber
\eea

\begin{Cor}\label{C8} If $\nabla+s(\Box)$ is an n-Jacobian, then
$$
\Delta_o^{o}(f_1,...,f_{n-1})=0\quad\mbox{and}\quad\Delta_o^1(f_1,...,f_{n-1})=0
$$
\end{Cor}
{\it Proof.} Corollary $\ref{C7}$ shows that $ X_i(\Box)=0$ and $X_i(\nabla)=0$.
Also we have
\bea
(-1)^{i-1}f_i\rfloor X_i=
(-1)^{i-1}f_i\rfloor\Box_{f_1,...,f_{i-1},f_{i+1},...,f_{n-1}}=\nonumber\\
(-1)^{i-1}\Box(f_1,...,f_{i-1},f_{i+1},...,f_{n-1},...,f_i)=
(-1)^{n-1}\Box(f_1,...,f_{n-1})=h\nonumber
\eea
Hence,
$$
-\sum_{i=1}^{n-1}(-1)^{i-1} X_i\wedge\nabla_{f_i}+(1-h)\nabla=
\sum_{i=1}^{n-1}(-1)^{i-1}f_i\rfloor (X_i\wedge\nabla)
$$
and
$$
-\sum_{i=1}^{n-1}(-1)^{i-1} X_i\wedge\Box_{f_i}+(1-h)\Box=
\sum_{i=1}^{n-1}(-1)^{i-1}f_i\rfloor(X_i\wedge\Box)
\qquad\triangleright
$$

\begin{Prop}\label{P14} 
($n-1$)-differential operators 
$\Delta_o^{o}$ and $\Delta_o^1$
satisfies relations
\bea\label{23}
\Delta_o^{o}(\varphi\psi,g_1,...,g_{n-2})=
\varphi\Delta_o^{o}(\psi,g_1,...,g_{n-2})+
\psi\Delta_o^{o}(\varphi,g_1,...,g_{n-2})\nonumber\\
-\nabla_{\varphi,g_1,...,g_{n-2}}\wedge\Box_\psi-
\nabla_{\psi,g_1,...,g_{n-2}}\wedge\Box_{\varphi}-\nonumber\\
(-1)^{n-1}\Box(\varphi,g_1,...,g_{n-2})\nabla_{\psi}-
(-1)^{n-1}\Box(\psi,g_1,...,g_{n-2})\nabla_\varphi
\eea
and
\bea\label{24}
\Delta_o^1(\varphi\psi,g_1,...,g_{n-2})=
\varphi\Delta_o^1(\psi,g_1,...,g_{n-2})+
\psi\Delta_o^1(\varphi,g_1,...,g_{n-2})\nonumber\\
-\nabla_{\varphi,g_1,...,g_{n-2}}\wedge\nabla_\psi-
\nabla_{\psi,g_1,...,g_{n-2}}\wedge\nabla_\varphi
\eea
\end{Prop}
{\it Proof.} This is essentially the 
same as the proof of proposition $\ref{P8}$.
One has to make use of the fact that the maps 
$f\longmapsto\nabla_f$ and $f\longmapsto\Box_f$
are derivations and to apply lemma $\ref{L4}$.$\qquad\triangleright$

\begin{Cor}\label{C9} If~ $\nabla+s(\Box)$ is $n$-Jacobian, then 
\bea\label{25}
\nabla_{\varphi,g_1,...,g_{n-2}}\wedge\Box_\psi
&+&\nabla_{\psi,g_1,...,g_{n-2}}\wedge\Box_\varphi\cr
+(-1)^{n-1}[\Box(\varphi,g_1,...,g_{n-2})\nabla_\psi
&+&\Box(\psi,g_1,...,g_{n-2})\nabla_\varphi]=0
\eea
and 
\be\label{26}
\nabla_{\varphi,g_1,...,g_{n-2}}\wedge\nabla_\psi+
\nabla_{\psi,g_1,...,g_{n-2}}\wedge\nabla_\varphi=0
\ee
\end{Cor}
{\it Proof.} Immediately from formulae ($\ref{23}$) and 
($\ref{24}$) and 
corollary $\ref{C8}$.\quad $\triangleright$

\begin{Cor}\label{C10} If $\nabla+s(\Box)$ is $n$-Jacobian, then 
the $n$-vector, corresponding to
$\nabla$ is locally either of rank $n$ (i.e locally decomposable) 
for $n>2$, or trivial.
\end{Cor}
{\it Proof.} Observe that theorem $\ref{Th1}$ results from formula ($\ref{e18}$) 
which is identical to ($\ref{26}$).$\quad\triangleright$

Denote by $V$ and $W$ multi-vectors corresponding to $\nabla$ and $\Box$
 respectively. Let
$\Pi_x$ and $P_x,~~x\in M$ be subspaces of 
$T_x M$ generated  by derived
vectors of $V_x$ and $W_x$ respectively.

\begin{Prop}\label{P15} 
If $\nabla+s(\Box)$ is $n$-Jacobian 
with $n>2$, then $rank(W_x)\leq n-1$ and $P_x\subset \Pi_x$ if $V_x\neq 0$
\end{Prop}
{\it Proof.} Relation $\Box_{g_1,...,g_{n-2}}(\nabla)=0$ 
(corollary $\ref{C7}$) implies 
\be\label{27} 
\Box_{\varphi,g_1,...,g_{n-3}}\wedge\nabla_\psi+\Box_{\psi,g_1,...g_{1-3}}
\wedge\nabla_\varphi=0
\ee
This can be proved repeating literally the reasoning used above to deduce
formula ($\ref{e18}$).
In terms of multi-vectors relation  ($\ref{27}$) is equivalent to
\be
(dg_{n-3}\rfloor...\rfloor dg_1\rfloor d\varphi\rfloor W)
\wedge(d\psi\rfloor V)+
(dg_{n-3}\rfloor...\rfloor dg_1\rfloor d\psi\rfloor W)
\wedge(d\varphi\rfloor V)=0\nonumber
\ee
In particular, for $\varphi=\psi$ we have
\be\label{28}
(dg_{n-3}\rfloor...\rfloor dg_1\rfloor d\varphi\rfloor W)
\wedge(d\varphi\rfloor V)
\ee
By lemma $\ref{LC}$ ($\ref{28}$) shows that the derived vector 
$$
(dg_{n-3}\rfloor...\rfloor dg_1\rfloor d\varphi\rfloor W)
$$ 
divides $d\varphi\rfloor V$.
 Since $V$  is of rank $n$ it divides also $V$. This proves the 
inclusion $P_x\subset\Pi_x$.
\par
Further, being $W$~~($n-1$)-Poissonian (corollary $\ref{C7}$) $rank(W) 
\leq n-1$ if $n>3$. For $n=3$
the inclusion $P_x\subset\Pi_x$ shows that $rank(W)\leq 3$
 due to decomposability of  $V$. But the rank of a bi-vector 
is an even number. 
So, $rank(W)\leq 2.\qquad\triangleright$

\begin{Cor}\label{C11} If $\nabla+s(\Box)$ is n-Jacobian with $n>2$, then
 $X_i\wedge\nabla=0$ and $X_i\wedge\Box=0$.
\end{Cor}
{\it Proof.}  $X_i$ is a derived vector of $W$ and, due to 
inclusion $P_x\subset\Pi_x$,
is also a derived vector of $V$. It remains to observe that a 
decomposable multi-vector vanishes when 
being multiplied by  any of its derived vectors.$\quad\triangleright$      

\begin{Cor}\label{C12} If $\nabla+s(\Box)$ is $n$-Jacobian 
with $n>2$, then
\be\label{29}
\nabla_{f_1,...,f_{n-1}}(\nabla)=0
\ee
\be\label{30}
\nabla_{f_1,...,f_{n-1}}(\Box)=\nabla_h
\ee
In particular, $\nabla$ is an $n$-Poisson structure on $M$.
\end{Cor}
{\it Proof.} Immediately from corollary $\ref{C8}$.\quad $\triangleright$
\par
Below it is supposed that $\Delta=\nabla+s(\Box)$ defines an 
$n$-Jacobi structure on $M$ with 
$n>2$. A point $x\in M$ of that $n$-Poisson manifold is called 
{\it regular} if both 
multi-vectors $V$ and $W$ corresponding to $\nabla$ and $\Box$,
respectively, do not vanish
at $x$. Note that the inclusion  $ P_x\subset \Pi_x$ 
(proposition $\ref{P15}$) implies that 
$x$ is regular if  $\Box$ is regular at $x$ , i.e. 
${W}_x\ne 0$.

Now we can prove the main struc\-tural result con\-cern\-ing 
$n$-Jacobian manifolds with $n>2$.

\begin{Th}\label{Th2} 
Let $\Delta$ be a non-trivial  
$n$-Jacobi structure and $n>2$. 
Then in a neighborhood of
any of its regular points it is either of the form 
$\Delta=\nabla+s(\nabla_h)$ 
where $\nabla$
is a non-trivial $n$-Poisson structure, or $\Delta=s(\Box)$ 
where $\Box$ is an $(n-1)$-Poisson structure (of rank 2 if $n=3$).
\end{Th}
{\it Proof.} Corollary $\ref{C7}$ and proposition $\ref{P15}$ show that $\Box$ is an 
($n-1$)-Poisson 
structure of rank $\le n-1$ on $M$ while corollaries $\ref{C10}$ and 
$\ref{C12}$ show $\nabla$
to be an $n$-Poisson one of rank $n$. Hamiltonian foliations of these two 
multi-Poisson structures
(we call them  $\Box$-foliaton and  $\nabla$-foliation, respectively) 
are regular foliations
of dimensions $n-1$ and $n$, respectively, in a neighborhood of a regular point 
$a\in M$.
Moreover,  $\Box$-foliation is inscribed into $\nabla$-foliation 
according to proposition $\ref{P15}$.
So, if the neighborhood $\cal U$ of $a$ is sufficiently small 
there exist a system of 
functionally independent functions $y,z_1,...,z_{m-n}, m=dim M$ 
such that they all 
are constant along leaves of the $\Box$-foliation and $z_1,...,z_{m-n}$ 
are constant 
along leaves  of the $\nabla$-foliation.
\par
Since $\Box$ is ($n-1$)-Poisson of rank $n-1$ there exist (locally)
 mutually commuting vector fields $X_1,...,X_{n-1}$ such that
 $\Box= X_1\wedge,...,\wedge X_{n-1}$.
 We can assume that $X_i\in D({\cal U})$. Then it is easy 
to see that there exist
functions $x_1,...,x_{n-1}\in {\cal C}^\infty(\cal U)$ such that 
$X_i(x_j)=\delta_{ij}$. Vector fields $X_i$'s are, 
obviously, tangent to leaves of 
$\Box$- foliation and, therefore, 
$X_i(y)=X_i(z_j)=0,\quad\forall j$.
By construction functions  
$x_1,...,x_{n-1},y,z_1,...,z_{m-n}$ 
are independent (functionally). So they form a local 
chart in $\cal U$ in, maybe,
 smaller neighborhood of $a$. Now vector fields 
$X_i$'s are 
identified with $\partial\over{\partial x_i}$'s, partial 
derivations in the sense of the above 
local chart. Note also, that the vector field 
$\frac{\partial}{\partial y}$        
is tangent to leaves of $\nabla$-foliation.
By construction the $n$-vector $V$ is tangent 
also to this leaves.
By this reason 
$\nabla=\lambda\frac{\partial}{\partial y}
\wedge\frac{\partial}{\partial {x_1}}\wedge
,...,\wedge\frac{\partial}{\partial x_{n-1}}$ 
with $\lambda\in{\cal C}^\infty(\cal U)$.
\par
Observe now that $\frac{\partial}{\partial x_i}$ is 
a $\Box$-Hamiltonian vector field 
associated with the Hamil\-tonian 
$((-1)^{i-1}x_1,x_2,...,x_{i-1},x_{i+1},...,x_{n-1})$ .
\par
For this field relation
$\Box_{g_1,...,g_{n-2}}(\nabla)=0$ (corollary $\ref{C7}$) becomes
\be
\frac{\partial}{\partial x_i}(\lambda \frac{\partial}{\partial y}
\wedge\frac{\partial}{\partial x_1}
\wedge\frac{\partial}{\partial x_{n-1}})=0\nonumber
\ee
which is equivalent to $\frac{\partial \lambda}{\partial x_i}=0$. 
This shows that $\lambda=\lambda(y,z_1,,...,z_{m-n})$. 
Hence, vector fields  
$X_1=\frac{\partial}{\partial x_1},...,X_{n-1}=
\frac{\partial}{\partial x_{n-1}},
\quad X_n=\lambda\frac{\partial}{\partial y}$ commute and , 
therefore, there exist functions 
$y_1,...,y_n\in{\cal C}^\infty(\cal U)$
such that $X_i(y_i)=\delta_{ij},\quad i,j=1,...,n$. 
Obviously, functions
 $y_1,...,y_n,z_1,...,z_{m-n}$ constitute a local chart with 
respect to which 
$X_i=\frac{\partial}{\partial y_i}, i=1,...,n$. 
Thus, we have proved that
\be\label{A31}
\nabla=\frac{\partial}{\partial y_1}\wedge,...,\wedge
\frac{\partial}{\partial y_n}\quad,
\quad\Box=\frac{\partial}{\partial y_1}\wedge,...,\wedge
\frac{\partial}{\partial y_{n-1}}
\ee
It remains to note that $\Box=\nabla_h$ for $h=(-1)^{n-1}y_n$.
This proves the first part of the theorem.
\par
To prove the second one we observe that if $\nabla\equiv 0$ in the
canonical decomposition of $\Delta$, i.e. $\Delta=s(\Box)$,
corollaries $\ref{C7}$ and $\ref{C11}$  show that $\Box$ is an $(n-1)$-Poisson
structure of rank $n-1$. (In virtue of theorem $\ref{Th1}$ last condition is
essential only if $n=3$.) On the other hand, one can see
easily that when $\nabla\equiv 0$ any such Poisson structure satisfies
conditions $\Delta^1=0,\quad\Delta^o=0$ of 
proposition $\ref{P13}$.$\quad\triangleright$

\begin{Cor}\label{C13} If $M$ is an $n$-Jacobian manifold and $n>2$, 
then in a neighborhood of an its regular point a local chart 
$y_1,...,y_n,z_1,...,z_{m-n}$ exists
such that
$$
\{f_1,...,f_n\}=det \left\|{\partial f_{i}\over 
\partial y_{j}}\right\|+
\sum\limits_{k=1}^{n}(-1)^{k-1}f_k det \left\|
{\partial f_{i}\over \partial y_{j}}\right\|_k
$$
where $\left\|{\partial f_{i}\over \partial f_{j}}
\right\|_k$ is the $(n-1)\times (n-1)$ -
matrix obtained from the $n\times n$-matrix 
$\left\|{\partial f_{i}\over \partial y_{j}}\right\|$ 
by canceling its $k$-th row
and $n$-th column.
\end{Cor}
{\it Proof.} It results directly from ($\ref{A31}$) and the 
definition of $s$.$\quad\triangleright$

\begin{Prop}\label{P16} 
Let $\nabla$ be an $n$-Poisson 
structure of rank $n$
 on $M$ and $f\in {\cal C}^\infty(M)$.
Then $\Delta=\nabla+s(\nabla_f)$ is an $n$-Jacobi structure. 
In particular,
 this is the case for any $n$-Poisson $\nabla$ with $n>2$.
\end{Prop}
{\it Proof.} With the no\-ta\-tion of propo\-si\-tion $\ref{P13}$ 
 $X_i=\nabla_{f,f_1,...,f_{i-1},f_{i+1},...,f_{n-1}}$ and $\Box=\nabla_f$.
 By this reason
$X_i(\nabla)=0$ as well as $\nabla_{f_1,...,f_{n-1}}(\nabla)=0$.
\par
Therefore, the $n$-differential operator $\Delta^1(f_1,...,f_{n-1})$ 
(proposition $\ref{P13}$) is reduced to 
$\sum\limits_{i=1}^{n-1}(-1)^{i-1}f_i\rfloor(X_i\rfloor)\nabla$.
Moreover, $X_i\rfloor\nabla=0$ due to the fact that $\nabla$ 
is of rank $n$. 
Hence, in the considered context $\Delta^1(f_1,...,f_{n-1})=0$

Next, $X_i(\nabla_f)=0$ since $\nabla_f$ is an 
$(n-1)$-Poisson structure.
 
By applying formula ($\ref{e6}$)~ for $\delta=\nabla_{f_1,..,f_{n_1}}$
and $u=f$ we see that for
 $h=(-1)^{n-1}\Box(f_1,..,f_{n-1})=(-1)^{n-1}\nabla(f,f_1,..,f_{n-1})$
$$
\nabla_{f_1,...,f_{n-1}}(\nabla_f)-\nabla_h=
f\rfloor\nabla_{f_1,...,f_{n-1}}(\nabla)=0
$$
So, the ($n-1$)-differential operator $\Delta^{o}(f_1,...,f_{n-1})$ 
(proposition $\ref{P13}$) is reduced to 
$\sum\limits_{k=1}^{n-1}(-1)^{i-1}f_i\rfloor(X_i\wedge\nabla_f)$.
But $\nabla_f$ is obviously, of rank $\leq n-1$ and so 
$X_i\wedge\nabla_f=0$.
Hence, $\Delta^{o}(f_1,...,f_{n-1})=0$ 
which proves that $\Delta$ is $n$-Jacobian.$\quad\triangleright$

The construction of proposition $\ref{P16}$ can be generalized as follows. 
Let $\omega$ be a closed
differential form of order 1. For a multi-derivation $\nabla$ 
define another one $\nabla^\omega$ by putting locally 
$\nabla^{\omega}=\nabla_f$ if $\omega=df$.
This definition is, obviously, correct and allows to 
globalize proposition $\ref{P16}$.

\begin{Prop}\label{P17} 
If $\nabla$ is an $n$-Poisson structure 
of rank n, then $\Delta=\nabla+s(\nabla^\omega)$ is an $n$-Jacobi 
structure for any closed differential 1-form $\omega$.
\end{Prop}
{\it Proof.} It results directly from proposition $\ref{P16}$ and from the fact that 
the $n$-Jacobi identity for 
$\Delta$ is a multi-differential operator.$\quad\triangleright$

\begin{emexa}\label{E9} With notation of example $\ref{E7}$ consider the 
($n+1$)-Poisson structure 
$\nabla=\frac{\partial}{\partial {\theta}_1}\wedge,...,\wedge\frac{\partial}
{\partial{\theta}_{n+1}}$
on ($n+1$)-torus $T^{n+1}$. Then $\nabla^\omega$ with the closed 
but not exact on $T^{n+1}$  1-form $\omega=\alpha d{\theta}_1-d{\theta}_2$ 
gives the $n$-Poisson structure described in example $\ref{E7}$. 
Therefore the ($n-1$) -Jacobi structure 
$\Delta=\nabla+s(\nabla^\omega)$ on        
$T^{n+1}$ is such that the leaves of its {\it $\Box$-foliation}  
are everywhere dense
in the unique leaf,  $T^{n+1}$, of its {\it $\nabla$-foliation}.
\end{emexa}
It is not difficult to show that any $n$-Jacobi with $n>2$ structure on 
an $n$-dimensional 
manifold is of the form  $\nabla+s(\nabla^\omega)$ for suitable 
closed 1-form $\omega$ and $n$-Poisson structure $\nabla$ on $M$.             

\section{$n$-Bianchi classification}
In view of the conjecture of sect. 3 on the structure of $n$-Lie
algebras for $n>2$ a classification of $(n+1)$-dimensional $n$-Lie
algebras turns out to be of a particular interest. Such a classification, 
an analogue of that of Bianchi for 3-dimensional Lie algebras, is,
 in fact, already done in \cite{Fi85} by a direct algebraic approach. 
Below we get it in a transparent geometric way which, in addition, 
reveals some interesting peculiarities.
\par
To start with, observe that on an orientable $(n+1)$-dimensional 
manifold $M$ any $n$-vector $P$ can be given in the form
$$
P=\alpha\rfloor V
$$
with an 1-form $\alpha=\alpha_{P,V}$ and a (prescribed) volume 
$(n+1)$-vector field  $V$ on $M$, respectively. 
Obviously, 
$\alpha \rfloor P=0$. This means that $\alpha$ vanishes on the
$n$-dimensional distribution defined by $P$. 
\par
If $P$ is an $n$-Poisson one, this distribution is tangent to the corresponding
Hamiltonian foliation and as such is integrable. Therefore,
 $\alpha\wedge d\alpha=0$. In virtue of proposition $\ref{P6}$ this
 condition is sufficient for $P$ to be an $n$-Poisson vector
 field.

Let us call an $n$-Poisson structure {\it unimodular} with respect to
$V$ if, for any $n$-Hamiltonian vector field $X$, $L_X(V)=0$ 

\begin{Prop}\label{PX} 
An $n $-Poisson structure $P$ is $V$-unimodular iff 
$d\alpha_{P,V}=0$
\end{Prop}
{\it Proof.} Recall the general formula
\be\label{Lie}
 L_X(\alpha\rfloor V)=
\alpha \rfloor L_X(V)- L_X(\alpha)\rfloor V  
\ee
 which holds for arbitrary vector field $X$, differential form $\alpha$ and 
multi-vector field $V$. If $X$ is a $P$-Hamiltonian field with 
$P=\alpha\rfloor V$, then $L_X(\alpha\rfloor V)=0$ and
($\ref{Lie}$) gives
$$
\alpha\rfloor L_X(V)= L_X(\alpha)\rfloor V
$$
Since, also, $X\rfloor\alpha=0\quad L_X(\alpha)= X\rfloor d\alpha$
 and the last equality can be rewritten as
\be\label{DIV}
div_{V} X \cdot P=(X\rfloor d\alpha)\rfloor V
\ee
due to the fact that $L_X(V)= div_VX\cdot V$.
So, $div_V X =0 \Leftrightarrow  L_X(V)=0$ for
any $P$-Hamiltonian field $X$ if $d\alpha=0$.
\par
Conversely, ($\ref{DIV}$) shows that $X\rfloor d\alpha$ vanishes
for any $P$-Hamiltonian field $X$ if $P$ is $V$-unimodular.
This implies that $Y\rfloor d\alpha=0$ for any $Y$ tangent to the 
hamiltonian foliation of $P$. Since this foliation is of codimension 1
any decomposable bi-vector $B$ on $M$ can be presented at least
locally, in form $B=Z \wedge Y$ with $Y$ as above. This shows that 
$BJ\rfloor d\alpha=0$ for any decomposable $B$ and, hence, $d\alpha=0.\qquad\triangleright$ 

Now we specify the above construction to the case $M={\cal V}^*, {\cal V}$ being an
 $(n+1)$-dimensional vector space
 and $P=T$, $T$ being the $n$-Poisson structure on ${\cal V}^*$ associated with 
an $n$-Lie algebra structure on ${\cal V}$. 
Also, we consider the $(n+1)$-vector field 
$V=\frac{\partial}{\partial x_1}\wedge\dots
\wedge\frac{\partial}{\partial x_{n+1}}$ on ${\cal V}$ where $x_i$'s
 are some cartesian 
coordinates on ${\cal V}^*$. Such an $(n+1)$-field is defined uniquely 
up to a scalar factor. So, the above concept of unimodularity does 
not depend on the choice of such a $V$ and the 1-form
$\alpha_{T,V}$ is defined uniquely up to a scalar factor.
Note also that $\alpha_{T,\cal C}$ is {\it linear} in the sense 
that the function $\Xi \rfloor \alpha_{T,\cal C}$ is linear
on ${\cal V}^*$, i.e. an element of ${\cal V}$, for any constant 
vector field $\Xi$.
 In coordinates this means that $\alpha_{T,V}$ looks as
$$
\alpha_{T,V}= \sum_{i,j}a_{ij}x_{j}dx_{i},\quad a_{ij}\in \bf R
$$

\begin{Prop}\label{PX1} 
Algebraic variety of $n$-Lie algebra structures 
on ${\cal V}$ is identical to the variety of linear differential 1-forms 
on ${\cal V}^*$ satisfying the condition $\alpha\wedge d\alpha=0$.
\end{Prop}
{\it Proof.} It was already shown that any $n$-Lie algebra structure 
on ${\cal V}$ is characterized uniquely by the corresponding linear 
differential 1-form
$\alpha_{T,\cal C}$. 
\par
Conversely, if $\alpha$ is a linear differential 1-form, 
then $n$-ary operation on ${\cal C}^\infty({\cal V}^*)$ defined by
$n$-vector field $\alpha\rfloor V$ is closed on the subspace
of linear functions on ${\cal V}^*$, i.e. on ${\cal V}$.
This way one gets an $n$-ary operation on $V$.
The condition $\alpha\wedge d\alpha=0$ guarantees integrability of 
the $n$-distribution on ${\cal V}^*$ defined by $P=\alpha\rfloor V$ 
and by virtue 
of the corollary $\ref{C4}$ it is an $n$-Poisson structure.
This fact restricted on $V$ shows the above $n$-ary
operation to be an $n$-Lie one. $\qquad\triangleright$
\par
Note now that any linear differential 1-form on ${\cal V}^*$ can be identified 
with a bilinear 2-form $b$ on ${\cal V}^*$. Namely, denote by $C_\omega$
the constant field of vectors on ${\cal V}^*$ which are equal to 
$\omega\in {\cal V}^*$ and put
$$
b(\omega,\rho) :=(C_{\omega}\rfloor \alpha,\rho),\quad\omega,\rho\in {\cal V}^*,
$$
where bracket $(\cdot,\cdot)$ stands for a natural pairing of ${\cal V}$ and
${\cal V}^*$. Obviously,
$$
b(\omega,\rho) =\sum_{i,j}a_{i,j}\omega_i\rho_j
$$
if $\omega=\sum \nolimits \omega_i\frac{\partial}{\partial x_i},\quad \rho=
\sum\nolimits \rho_j\frac{\partial}{\partial x_j}$ and $\alpha=
\sum\nolimits a_{ij}x_j\,dx_i$. So, $\Vert a_{ij}\Vert$ is the matrix of $b$.
The form $b$ is called {\it generating} for the $n$-Lie algebra in question.
\par
An $n$-Lie algebra is called {\it unimodular} if all its inner derivations are 
unimodular operators. For an $(n+1)$-dimensional $n$-Lie algebras this is,
obviously, equivalent to unimodularity of the associated Poisson structure
 $T$ on ${\cal V}^*$ with respect to a cartesian volume $(n+1)$-vector $V$.
On the other hand, $T$ is $V$-unimodular iff $d\alpha_{T,V}=0$
(proposition $\ref{PX}$) and for a linear differential 1-form $\alpha$ the 
condition $d\alpha=0$ is equivalent to $\alpha=dF$ for a bilinear polynomial 
$F$ on ${\cal V}^*$ (or to symmetry of the corresponding quadratic 
form $b)$.
These considerations prove the following result.

\begin{Prop}\label{PX2} 
The $n$-Poisson structure $T$ on ${\cal V}^*$ 
associated with an unimodular Lie algebra structure on an $(n+1)$
- dimensional vector space ${\cal V}$ is of the form $dF\rfloor{V}$
for a suitable quadratic polynomial $F$ on ${\cal V}^*$. Therefore, all 
unimodular $n$-Lie structures on ${\cal V}$ are mutually compatible.
Two such structures are isomorphic
iff the corresponding quadratic polynomials can be reduced one to another 
up to a scalar factor by a linear transformation. In particular, for $k= \bf R$
isomorphic classes of unimodular $(n+1)$-dimensional $n$-Lie structures 
can be labeled by two numbers: $r$ (the rank of $F$),\quad $0\leq r\leq n+1$
and $m$ (the maximal of positive and negative indices of $F$), 
$\frac{r}{2}\leq m\leq r$.$\quad\triangleright$
\end{Prop}
Passing now to the case $d\alpha_{T,V}\ne 0$ we note that
  $d\alpha_{T,V}$ is a constant differential 2-form on ${\cal V}^*$ 
due to linearity of  $\alpha_{T,V}$. Moreover, the condition 
 $\alpha_{T,V}\wedge d\alpha_{T,V}=0$ shows that the rank of 
 $d\alpha_{T,V}$ is equal to 2. Therefore,  
$d\alpha_{T,V}=dx_1\wedge dx_2$ in suitable cartesian coordinates 
on ${\cal V}^*$. Since $\alpha_{T,V}$ divides $dx_1\wedge dx_2$ 
and is linear it must be of the form 
$$
\sum_{i=1}^{2}\mu_{ij}x_{j}dx_i\quad\mbox{ with}\quad\mu_{21}-\mu_{12}=1.
$$ 
This is equivalent to say that $d\alpha_{T,V}=
dq+\frac{1}{2}(x_{1}dx_{2}-x_{2}dx_1)$ with 
$$
q=q(x_1,x_2)= \frac{1}{2}(\mu_{11}x_{1}^{2}+(\mu_{12}+
\mu_{21})x_{1}x_{2}+\mu_{22}x_{2}^{2}).
$$
Note that unimodular transformations of variables does
not alter the form of the skew-symmetric part of $\alpha_{T,V}$. 
So, by performing a suitable such one it is possible to 
reduce $q(x_{1},x_{2})$ to a diagonal form:
$$
\alpha_{T,V}=d(\mu y_{1}^{2}+\nu y_{2}^{2})+\frac{1}{2}(y_{1}dy_{2}-y_{2}dy_{1})
$$
Further, transformations of the form $(y_{1}, y_{2})\rightarrow
 (\lambda y_{1}, \pm\lambda^{-1}y_2)$ and the possibility to change
the sign of $\alpha_{T,V}$ allows to bring it to one of the following
canonical forms
\bea\label{LST}
&~&\Psi_{\lambda}^{\pm}(n):\frac{\lambda}{2}d(z_{1}^{2}\pm z_{2}^{2})+
\frac{1}{2}(z_{1}dz-z_{2}dz_1)\quad,\quad\lambda>0\cr
&~&\Psi_{1}(n):z_1dz_1+\frac{1}{2}(z_{1}dz_2-z_{2}dz_1),\cr
&~&\Psi(n):\frac{1}{2}(z_{1}dz_2-z_{2}dz_1)
\eea

\begin{Prop}\label{PX3} 
$n$-Lie algebras corresponding to the 1-form 
$\alpha_{T,V}$ of the list ($\ref{LST}$) are mutually non isomorphic
 and, therefore, label isomorphic classes of non-unimodular $(n+1)$
-dimensional $n$-Lie algebras.
\end{Prop}
{\it Proof.}~~~Previous considerations show that any non-unimodular
$(n+1)$-dim\-ensio\-nal $n$-Lie algebra is isomorphic to one of the list 
($\ref{LST}$). Two algebras of the type $\Psi_{\lambda}^{\pm}(n)$
corresponding to different $\lambda$ are not isomorphic since non-vanishing
of the skew-symmetric part of $\alpha_{T,V}$ is equivalent of 
non-unimodularity condition. On the other hand, $\lambda$ is a an 
invariant of isomorphism type since $\frac{2}{\lambda}$ is equal to the
area of a (quasi-) orthonormal base of the symmetric part of 
$\alpha_{T,V}$ measured by means of its skew-symmetric part.
Other types differ by rank or signature of the symmetric part.$\qquad\triangleright$
\par
The classification we have got has an interesting {\it internal} structure. 
Na\-me\-ly, denote by $B(n)$ the isomorphism type of $(n+1)$-dimensional 
$n$-Lie algebras corresponding to the generating polynomial 
$\frac{1}{2}x_{1}^2$.
Then any $(n+1)$-dimensional algebra can be seen as a "molecule" composed of
$B(n)$ and $\Psi(n)$ types of "atoms". More exactly, the above discussion 
can be resumed as follows
 
\begin{Prop}\label{Pre-Egregium}
Any $(n+1)$-dimensional $n$-Lie algebra 
can be realized as the sum of mutually compatible algebras each of them being
either of type $B(n)$ or of type $\Psi(n)$.
\end{Prop}
On the base of the obtained classification it is not difficult to describe
completely the derivation algebras of $(n+1)$-dimensional n-Lie algebras.

An linear operator $A:W\rightarrow W$ is called an infinitesimal conformal symmetry 
of a bilinear form $b(u,v)$ on $W$ if 
\bea\label{CS}
b(Au,v)+b(u,Av)=tr(A)b(u,v)
\eea  

\begin{Prop}\label{DER} 
The Lie al\-ge\-bra of deri\-vations of an 
$(n+1)$-dimensional $n$-Lie algebra coincides with the algebra
of infinitesimal conformal symmetries of its generating bilinear form.
\end{Prop}
{\it Proof.} A linear operator $A$ on a linear space can be naturally 
interpreted as
a linear vector field $X$ on it. Moreover, $tr(A)=div(X)$. Formula 
($\ref{Lie}$) for such
a field $X$ which is also a symmetry of $\alpha\rfloor V$ reduces to
$$
\alpha \rfloor L_X(V)=L_X(\alpha)\rfloor V  
$$
which is identical to ($\ref{CS}$).$\qquad\triangleright$

We omit a complete description of the derivation algebras which can be easily
got by applying the previous proposition. Just note that inner
derivations exhaust all derivations of an $(n+1)$-dimensional algebra
iff the rank of its generating
 form is equal to $n+1$. The following example 
illustrate some features of outer derivations. 

\begin{emexa}\label{E9+1}
Consider the 4-dimensional 3-Lie algebra corresponding to the generating polynomial
 $F={1\over 2}{x_4}^2$. The associated  3-Poisson tensor is
$$
P=x_4
\frac{\partial}{\partial x_1}\wedge 
\frac{\partial}{\partial x_2}\wedge 
\frac{\partial}{\partial x_3}
$$
Clearly fields
$
x_4\frac{\partial}{\partial x_1},~~x_4\frac{\partial}{\partial x_2},
~~x_4\frac{\partial}{\partial x_3}
$
form a basis of inner derivations.
\par
Proposition $\ref{DER}$ shows that
$$
x_4\frac{\partial}{\partial x_4}+ x_1\frac{\partial}{\partial x_1},
~~x_4\frac{\partial}{\partial x_4}+ x_2\frac{\partial}{\partial x_2},
~~x_4\frac{\partial}{\partial x_4}+ x_3\frac{\partial}{\partial x_3}         
$$
are outer derivations not tangent to the Hamiltonian leaves of P.
On the other hand,  the following outer derivations
$$
x_1\frac{\partial}{\partial x_1}-x_2\frac{\partial}{\partial x_2},
~~Jx_2\frac{\partial}{\partial x_2}-x_3\frac{\partial}{\partial x_3},
~~x_3\frac{\partial}{\partial x_3}-x_1\frac{\partial}{\partial x_1}         
$$
are tangent to these leaves.
\end{emexa}
\par
Previous method used to get the {\it $n$-Bianchi classification} can
be extended to inscribe into the $n$-ary context infinite dimensional
Lie algebras too. This is well illustrated by the following example.
\begin{emexa}{{\rm (Witt algebra)}}
The Witt (or $sl(2,{\bf R})$ Kac-Moody) algebra is generated by $e_i$
,~~$i\in (0,1,2,\cdots)$according to
$$
[e_i,e_j]=(j-i)e_{i+j-1},~~~\forall i,j\in {\bf N}
$$
Elements $e_0,e_1,e_2$ generate a 3-dimensional subalgebra isomorphic
to $sl(2,{\bf R})$. It is easy to see that the multiple commutator 
$\epsilon_k=\underbrace{[e_2,\cdots,[e_2,e_3]]}_{k~~times}$ is equal 
to $k!e_{3+k}$. So the
elements $e_0,e_1,e_2, e_3$ and $\epsilon_k, \forall k\in {\bf N}$
constitute a new basis of the Witt algebra.
\par
Let us consider now the Poisson bracket on ${\bf R}^3$ given by $P_F$
with
$$
P=\frac{\partial}{\partial x_1}\wedge 
\frac{\partial}{\partial x_2}\wedge 
\frac{\partial}{\partial x_3}
$$
and
$$
F=x_1x_3-x_2^2,
$$
i.e
$$
P_F=x_1\frac{\partial}{\partial x_1}\wedge 
\frac{\partial}{\partial x_2}+2x_2\frac{\partial}{\partial x_1}\wedge 
\frac{\partial}{\partial x_3}+x_3\frac{\partial}{\partial x_2}\wedge 
\frac{\partial}{\partial x_3}
$$

Then we have the following ordinary Poisson bracket:
$$
\{x_1,x_2\}=x_1,~\{x_1,x_3\}=2x_2,~J\{x_2,x_3\}=x_3
$$

So the correspondence:
\begin{equation}
[\cdot~,~\cdot]\leftrightarrow \{\cdot,\cdot\},
~e_0\leftrightarrow x_1,
~e_1\leftrightarrow x_2, 
~e_2\leftrightarrow x_3\nonumber
\end{equation}
is an isomorphism of Lie algebras. Moreover this isomorphism 
of subalgebras can be extended to an embedding of the whole 
Witt algebra into the Poisson algebra $\{\cdot,\cdot\}$ 
according to:
\begin{eqnarray}
[\cdot~,~\cdot]&\leftrightarrow& \{\cdot,\cdot\},\nonumber\cr
e_0&\leftrightarrow &x_1,\nonumber\cr
e_1&\leftrightarrow &x_2,\nonumber\cr 
e_2&\leftrightarrow &x_3,\nonumber\cr
e_3&\leftrightarrow &g={x_3^2\over x_2}
({2F\over x_1x_3}-{x_1x_3\over F}-1),\nonumber\cr
e_{3+k}&\leftrightarrow &{1\over k!}
\underbrace{\{x_3,\cdots ,\{x_3,g\}\}}_{ k~~times}
\end{eqnarray}

\end{emexa}
\section{Dynamical aspects} 

  A Hamiltonian vector field $X_{H_1,..,H_{n-1}}$ associated with 
an $n$-Poisson structure can be called {\it $n$-Poisson}, or {\it Nambu dynamics}. 
The corresponding equation of motion is
\be             
{df\over dt} =  X_{H_1, H_2,..., H_{n-1}}f= \{H_1, H_2,..., H_{n-1},f\}       
\ee
\par
An important peculiarity of a Nambu dynamics is that it admits at least $n-1$ independent 
constants of motion, namely $H_1,...,H_{n-1}$. Also such a dynamics admits $n-1$ different 
but mutually compatible Poisson descriptions. The corresponding $i$-th ({\it usual}) Poisson
bracket and Hamiltonian are
$$
\{f,g\}_i = \{H_1,...,H_{i-1},H_{i+1},...,H_{n-1},f,g\}~~~\mbox{and}~~~(-1)^{n-1}H_i, 
$$
respectively.
\par
So, the fact that a dynamics is a Nambu one can be exploited with the use. 
Below we give some examples of that.

\subsection{The Kepler dynamics}
Occasionally, a dynamical vector field $\Gamma$ admitting $2n-1$ constants
of the motion on a $2n$-dimensional manifold $M$, is called 
{\it hyper-integrable} or {\it degenerate}.
\par
In these cases denoting with $L_\Gamma$ the Lie derivative with respect to 
$\Gamma$ and with 

If $f_1, f_2,...,f_{2n-1}$ are first integrals for $\Gamma$ and
$f_{2n}\in C^{\infty}(M)$ is such that $\Gamma (f_{2n})=1$, then 
the $2n$-Poisson bracket
\be\label{KN}
\{ h_1, h_2,..., h_{2n}\} = 
det \left\|{\partial h_{i}\over 
\partial f_{j}}\right\|,~~~i,j\in (1,\cdots ,2n)\nonumber
\ee
is preserved by $\Gamma$ which becomes Hamiltonian with respect 
to ($\ref{KN}$) with the Hamiltonian function ($f_1, f_2,...,f_{2n-1}$).
\par
Of course the corresponding $2n$-Poisson vector is:
\be
\Lambda= {\partial\over\partial f_1}\wedge{\partial\over\partial f_2}\cdots
\wedge {\partial\over\partial f_{2n}}\nonumber
\ee
\par
More generally $2n$-Poisson bracket 
\be\label{NBI}
\{ h_1, h_2,..., h_{2n}\}_F = 
Fdet \left\|{\partial h_{i}\over 
\partial f_{j}}\right\|,~~~i,j\in 1,\cdots ,2n
\ee
is preserved by $\Gamma$
iff $F$ is a first integral, i.e. $F=F(f_1,f_2,...,f_{2n-1})$
\par
The Kepler dynamics illustrates such a situation
\par
Recall that the Kepler vector field, in spherical-polar 
co\-or\-di\-na\-tes ($r, \theta, \varphi$) in $R^3 - \{0\}$ , is given by:
\bea
\Gamma &=& {1\over m}(p_r{\partial\over \partial r}+
{p_\theta \over r^2}{\partial\over \partial\theta}+
{p_\varphi\over r^2sin^2\theta}{\partial\over \partial \varphi}\cr
&-&{1\over r^3}{(p_\theta^2 + p_\varphi^2)\over sin^2\theta}
{\partial\over \partial p_r}- 
{p_\varphi^2cos\theta \over r^2sin^3\theta}
{\partial\over \partial p_\theta}- 
{k\over r^2}{\partial\over \partial p_\varphi})  
\eea
with ($p_r, p_\theta, p_\varphi$) canonical conjugate variables.
\par 
$\Gamma$ is globally hamiltonian with respect to the symplectic form:

\begin{equation}		
\omega =  dp_r\wedge dr+dp_\theta\wedge d\theta+dp_\varphi\wedge d\varphi                                                             
\end{equation}

\noindent
with Hamiltonian $H$ given by (see, for instance \cite{LCA29}):

\begin{equation}		
H = {1\over 2m}(p_r^2 + {p_\theta^2\over r^2} + 
{p_\varphi^2\over r^2sin^2\theta} )-{k\over r} 
\end{equation}

In  action-angle coordinates $(J_h, \varphi_h),~~~~h\in (1,2,3)$ 
(see, for instance \cite{SC71}), 
the Kepler Hamiltonian $H$, the symplectic form  $\omega$ and the 
vector field  $\Gamma$  become:
	
\begin{eqnarray}\label{AAD}		
H &=& - {mk^2\over (J_r+J_\theta+J_\varphi)^2}\nonumber\\                                                       
\omega &=& \sum_hdJ_h\wedge d\varphi^h\nonumber\\                                                      
\Gamma &=& \nu
({\partial\over \partial \varphi_1}+
{\partial\over \partial \varphi_2}+
{\partial\over \partial \varphi_3})                   
\end{eqnarray}
with $\nu={2mk^2\over (J_r+J_\theta+J_\varphi)^3}$
\par	
Functionally independent constants of the motion are:
\par
$ f_1= J_1, f_2= J_2 , f_3= J_3, 
f_4= \varphi_1 - \varphi_2, f_5= \varphi_2 - \varphi_3 $
\par
Now it is easy to see that ($\ref{AAD}$) becomes 6-Hamiltonian
with respect to ($\ref{NBI}$) with $F=\nu$
\par
So
\be
\{ h_1, h_2, h_3, h_4, h_5, h_6\} = \nu {\partial (h_1, h_2, h_3, h_4, h_5, h_6)\over 
\partial (J_1 , J_2 , J_3, \varphi_1 , \varphi_2, \varphi_3)}
\ee
provides us with a $6$-ary bracket for the Kepler dynamics.

In terms of this  bracket, the equations of the motion looks as:
\be
{df\over dt}=\nu \{J_1 , J_2 , J_3, \varphi_1- \varphi_2, \varphi_2 - \varphi_3, f\}
\ee

By fixing same of the functions $h$'s we get hereditary brackets.

\subsection{The spinning particle}

Given a dynamics, i.e. a vector field $\Gamma$ on a manifold $M$, it colud be 
interesting to realize it as a Hamiltonian field with respect to a Poisson
structure \cite{CIMS95}. Below it wil be shown how multi-Poisson 
structures can be used in this connection.

\par
We shall ignore the spatial degree of freedom of the particle and study 
only the spin variables. Let us denote the spin variables ${\bf S}=S_1, S_2, S_3$
as elements in ${\bf R}^3$. The equations for these variables when the 
particle interacts with an external magnetic field ${\bf B}=B_1, B_2, B_3$ are given
by:
\be\label{SD}
{dS_i\over dt}=\mu\epsilon_{ijk}S_jB_k
\ee
where $\mu$ denotes the magnetic moment.
\par
This dynamics has two first integrals, namely, ${\bf S}^2=S_1^2+S_2^2+S_3^2$ and 
${\bf S}\cdot {\bf B}=S_1B_1+S_2B_2+S_3B_3$ and,
in addition, is canonical for the ternary bracket 
associated with the 3-vector field 
$$
{\partial\over \partial S_1}\wedge {\partial\over \partial S_2}\wedge 
{\partial\over \partial S_3}
$$

The most general ternary bracket preserved by dynamics ($\ref{SD}$),
is associated with the three vector field
\be\label{spb}
f{\partial\over \partial S_1}\wedge {\partial\over \partial S_2}\wedge 
{\partial\over \partial S_3}
\ee
where $f$ is a first integral of it.
\par
All Poisson structures obtained by fixing a function 
$F=F(S^2,{\bf S}\cdot{\bf B})$, are preserved by the dynamics 
and are mutually compatible. The corresponding Poisson bracket
is:
$$
\{S_j,S_k\}^f_F=f\epsilon_{jkl}{\partial F\over \partial S_l}
$$

Now we show how the ternary Poisson structure ($\ref{spb}$) 
allows for the alternative ordinary Poisson brackets described
in \cite{CIMS95}:

\begin{itemize}
\item
{\sl Standard description}
$$
f={1\over 2}, F= S^2 
$$

For this choice the algebra generated by the Poisson brackets on
linear functions is the $su(2)$ Lie algebra. The Hamiltonian
function for the dynamics is the standard one $H=-\mu{\bf S}\dot{\bf B}$.
\item
{\sl Non-standard description}

Now we take 
$$
f={1\over 2}, F= S_1^2+S_2^2+
{1\over 2\lambda}[{cosh2\lambda S_3\over sinh\lambda}-{1\over \lambda}] 
$$
with Hamiltonian $H=-\mu\lambda S_3$. Here for simplicity
we have taken the magnetic field along the third axis. The parameter
$\lambda$ is a deformation parameter and the standard description
is recovered for $\lambda\mapsto 0$.
\par
The hereditary Poisson brackets are:
\bea\nn
&\{S_2,S_3\}_F^f= S_1\cr
&\{S_1,S_3\}_F^f= S_2\cr
&\{S_1,S_2\}_F^f= {1\over 2}{sinh2\lambda S_3\over sinh\lambda}
\eea
\par
These brackets are a classical realization of the quantum
commutation relations for generators of the $U_q(sl(2))$
Hopf algebra.

We also notice that this Poisson Bracket is compatible with the
previous one as they are hereditary from the same ternary structure
($\ref{spb}$).
\item
{\sl Another non-standard description}

There is another choice for $f$ and $F$ which is known to 
correspond to the classical limit of the $U_q(sl(2))$
Hopf algebra.
\par
It is 
$$
f={\lambda\over 4}S_3,  F= S_1^2+S_2^2+S_3^2+S_3^{-2}
$$
It leads to the following brackets:
\bea\nn
&\{S_2,S_3\}^f_F= {\lambda\over 2}S_1S_3\cr
&\{S_1,S_3\}^f_F= \lambda S_2S_3\cr
&\{S_1,S_2\}^f_F= {\lambda\over 2}[S_3^2-S_3^{-2}]
\eea
With respect to this Poisson bracket dynamics ($\ref{SD}$) becomes
Hamiltonian  with Hamiltonian function:
$$
H=-{2\mu B\over \lambda}ln S_3
$$
with the magnetic field along the third axis. 
\end{itemize}

Of course dynamics ($\ref{SD}$) admits many other Poisson realizations
of this type.
 

\newpage

\begin{thebibliography}{99}

\bibitem{AG96}
D. Alexeevsky and Guha, {\it On Decomposability of Nambu-Poisson Tensor}
Acta Mathematica Universitatis Comenianae, Vol. {\bf 65} (1996)1-9 
\bibitem{CV92}
A.Cabras, A.M.Vinogradov, {\it Extensions of the Poisson bracket 
to differential forms and multivector fields}, J.Geom.Phys. {\bf 9}
(1992),75-100.
\bibitem{CIMP94}
J.F.Cari$\tilde{n}$ena, L.A.Ibort, G.Marmo, A.Perelomov,
{\it The Geometry of Poisson manifolds and Lie Algebras}
J.Phys. A:  Math and Gen  {\bf 27}, 7425  1994
\bibitem{CIMS95}
J.F.Cari$\tilde{n}$ena, L.A.Ibort, G.Marmo, A.Stern
{\it The Feynman problem and the inverse problem for Poisson
dynamics} Physics Reports {\bf 263} (1995)
\bibitem{DMSV82} 
S. De Filippo, G. Marmo, M. Salerno, G. Vilasi,
{\it On the Phase Manifold Geometry of Integrable Nonlinear  Field 
Theory}, Preprint IFUSA, Salerno (1982), unpublished. 
\bibitem{DMSV84} 
S. De Filippo, G. Marmo, M. Salerno, G. Vilasi,
{\it A  New Characterization of Completely Integrable Systems.}
\NCB {\bf 83}, 97 (1984) 
\bibitem{Fi85}
V.T.Filippov, {\it n-Lie Algebras}, Sibirskii Mathematicheskii Zhurnal {\bf 26}, n.6
126 (1985) 
\bibitem{FN56}
A.Frolicher and A.Nijenhuis,{\it Theory of vector valued differential forms I} 
\IM {\bf 23}, 338 (1956)
\bibitem{Gned}A.V.Gnedbaye {\it Les alg\`ebres $k$-aires et leur op\`erades}
C. R. Acad. Sci. Paris, S\`erie I {\bf 321}(1995)
\bibitem{GMP93}J.Grabowski,
{\it Abstract Jacobi and Poisson structures. Quantization and
star-products}
Jour.Geom.Phys. {\bf 9} (1992) 45-73
\bibitem{Ki76}A.A.Kirillov, {\it Local Lie Algebras} Usptkhi Mat.Nauk {\bf 31}:4
(1976)57-76; Russian Math Surveys {\bf 31}:4 (1976)55-76  
\bibitem{KLV}I. S. Krassil'shchik, V. V. Lychagin, A. M. Vinogradov 
{\it Ge\-om\-e\-try of jet spaces and nonlinear partial differential equations}
Gordon and Breach, N.Y. 1986  
\bibitem{LCA29} T.Levi-Civita e U.Amaldi, {\it Lezioni di Meccanica Razionale}
Zanichelli  (Bologna 1929)
\bibitem{LMV94} 
G.Landi, G.Marmo and G.Vilasi,{\it Recursion Operators: Meaning and Existence}
 \JMP{\bf 35}, n.2 808 (1994)
\bibitem{Loday}J.L.Loday, {\it La renaissance des operades}, Sem. Bourbaki
47\`eme annee, 1994-95, n! 792.
\bibitem{Ly}V.V.Lychagin, {\it A local classification of non-linear
first order partial differential equations} Usptkhi Mat.Nauk {\bf 30}:1
(1975)101-171; Russian Math Surveys {\bf 30}:1 (1975)105-175
\bibitem{Ma78} 
F.Magri, {\it A simple model of in\-te\-gra\-ble Hamil\-tonian equa\-tion} 
 \JMP  {\bf 19}, 1156 (1978) 
\bibitem{Ma80} 
F.Magri,{\it A Geometrical Approach to the Nonlinear Solvable Equations} 
Lect. Notes in Phys. {\bf 120} 233 (1980) . 
\bibitem{Mich Vin}
P. Michor and A. M. Vinogradov, {\it $n$-ary Lie and as\-sociative algebras},
ESI preprint, December 1996, to appear in Proceedings of the Conference
{\it Geometry and Physics} Vietri sul Mare, October 1996.  
\bibitem{Na73}
Y.Nambu,{\it Generalized Hamiltonian Mechanics} \PRD {\bf 7}, 2405 (1973)
\bibitem{Ni87}
A.Nijenhuis, {\it Trace-free differential invariants of triples of vector
1-forms} \IM {\bf 49},2 (1987).
\bibitem{SC71} E.J.Saletan and A.H.Cromer, {\it Theoretical Mechanics}
J.Wiley \& Sons (N.Y 1971)
\bibitem{Ta94}L.A.Takhtajan,{\it On Foundation of Generalized Nambu Mechanics} 
\CMP {\bf 160}, 295 (1994)
\bibitem{Vin}
A.M.Vinogradov {\it The logic algebra for the theory of linear differential operators
} Sov. Math. Dokl. {\bf 13}(1972), 1058-1062
\bibitem{Vin-Kra}
A.M.Vinogradov, I. S. Krassil'shchik, {\it What is the Hamiltonian formalism},
Russian Math. Surveys vol. 30(1975)177-202.
\bibitem{VinC}
A.M.Vinogradov, {\it The $C$ spectral sequence, Lagrangian 
formalism and conservation laws:I The linear theory: II The non-linear theory.}
J.Math. Anal. and Appl. {\bf 100}(1984), 1-40, 41-129.
\bibitem{VV}
A.Vinogradov and A.M.Vinogradov, {\it Alternative $n$-Poisson manifolds}~~in progress
\end{thebibliography}
\end{document}